\documentclass[
    aps,
    prd,
    reprint,
    superscriptaddress,
    nolongbibliography,
    nobibnotes,
    nofootinbib,
]{revtex4-2}

\usepackage[dvipsnames, usenames]{xcolor}
\definecolor{linkcolor}{rgb}{0.1, 0.5, 0.7}

\usepackage[
colorlinks=true,
linkcolor=linkcolor,
citecolor=linkcolor,
filecolor=linkcolor,
urlcolor=linkcolor,
pdfusetitle,
]{hyperref}

\usepackage{amssymb}
\usepackage{amsmath}
\usepackage{physics}
\usepackage{orcidlink}
\usepackage{acronym}
\usepackage{booktabs}

\usepackage[caption=false]{subfig}

\usepackage[normalem]{ulem}
\usepackage[T1]{fontenc}

\newcommand{\comment}[1]{}

\newcommand{\ligo}{\affiliation{LIGO Laboratory, Massachusetts Institute of Technology, Cambridge, MA 02139, USA}}
\newcommand{\mki}{\affiliation{Kavli Institute for Astrophysics and Space Research, Massachusetts Institute of Technology, Cambridge, MA 02139, USA}}
\renewcommand{\mit}{\affiliation{Department of Physics, Massachusetts Institute of Technology, Cambridge, MA 02139, USA}}
\newcommand{\notts}{\affiliation{Nottingham Centre of Gravity \& School of Mathematical Sciences, University of Nottingham,\\University Park, Nottingham, NG7 2RD, United Kingdom}}
\newcommand{\gla}
{\affiliation{Institute for Gravitational Research, School of Physics and Astronomy, University of Glasgow, Glasgow, G12 8QQ, United Kingdom}}

\newacro{LVK}[LVK]{LIGO--Virgo--KAGRA}

\newacro{BBH}{binary black hole}
\newacro{GW}{gravitational wave}
\newacro{SNR}{signal-to-noise ratio}
\newacro{FAR}{false-alarm rate}
\newacro{KL}{Kullback--Leibler}
\newacro{HMC}{Hamiltonian Monte Carlo}
\newacro{VI}{variational inference}
\newacro{PSIS}{Pareto-smoothed importance sampling}
\newacro{ICAR}{intrinsic conditional autoregressive}
\newacro{PPD}{posterior population distribution}
\newacro{XG}{next-generation}
\newacro{BH}{black hole}
\newacro{GWTC}{gravitational-wave transient catalog}

\graphicspath {{./figures/}}

\def\var{\ensuremath{\mathcal{V}}}
\def\cat{\ensuremath{\mathcal{D}}}
\def\src{\ensuremath{\Theta}}
\def\rnorm{\ensuremath{K}}

\begin{document}

\title{Neural Bayesian updates to populations with growing gravitational-wave catalogs}

\author{Noah E. Wolfe\,\orcidlink{0000-0003-2540-3845}}
\email{newolfe@mit.edu}
\ligo\mki\mit
\author{Matthew Mould\,\orcidlink{0000-0001-5460-2910}}
\notts\ligo\mki\mit
\author{John Veitch\,\orcidlink{0000-0002-6508-0713}}
\gla
\author{Salvatore Vitale\,\orcidlink{0000-0003-2700-0767}}
\ligo\mki\mit

\date{\today}

\begin{abstract}
As gravitational-wave catalogs grow, they will become increasingly computationally expensive to analyze in their entirety, especially when inferring astrophysical source populations with high-dimensional, flexible models.
Bayesian statistics offers a natural remedy,
letting us update our knowledge of physical models as new data arrive,
without re-analyzing existing data.
However, doing so requires the posterior probability density of model parameters
for previous observations, which is typically intractable.
Here, we use variational neural posterior estimation
to rapidly update the inferred population of binary black holes as data are observed in gravitational-wave detectors.
We apply this approach to real and simulated catalogs analyzed with both low- and high-dimensional population models,
testing the reliability of three update cadences:
with new catalogs of sources,
month by month during an observing run,
and as each new signal arrives.
We investigate the success and failure modes of neural sequential updates, finding that the robustness of updating is sensitive to the information contained in each update
and that updating is most effective when performed with larger segments of data.
We outline one additional scientific application enabled by Bayesian updating: identification of events that are individually informative about the population.
Neural Bayesian updates to astrophysical population models also provide efficient likelihood representations for joint analyses with other data, e.g., standard-siren cosmology,
and similar methods can be used to perform Bayesian stochastic background searches.
\end{abstract}

\maketitle

\section{Introduction}

The \ac{LVK} \cite{LIGOScientific:2014pky, VIRGO:2014yos, KAGRA:2020tym} catalog of \ac{GW} events has grown to 204 sources from the first detection in 2015 \cite{LIGOScientific:2025slb}.
The number of observed \acp{BBH} in particular is expected to grow to thousands by the end of the \ac{LVK}'s next observing run \cite{Kiendrebeogo:2023hzf}, and tens to hundreds of thousands when next-generation detectors come online \cite{Gupta:2023lga}.
Simultaneously, population models have grown in dimension and complexity
as analysts search for ever-more
subtle features \cite{Farah:2023vsc, Galaudage:2024meo, Biscoveanu:2025jpc, 2021ApJ...913L..19T, Tiwari:2025lit, Willcox:2025cus},
subpopulations \cite{LIGOScientific:2025pvj, Vijaykumar:2026zjy, Farah:2026jlc, Mould:2022ccw, Li:2023yyt, Godfrey:2023oxb, Hussain:2024qzl, Sadiq:2025vly, Banagiri:2025dmy, Adamcewicz:2025phm, Sridhar:2025kvi, Antonini:2025ilj, Tong:2025xir, Tong:2025wpz, Plunkett:2026pxt, Stegmann:2025zkb}
and correlations \cite{LIGOScientific:2025pvj, Vijaykumar:2026zjy, Farah:2026jlc, Callister:2021fpo, Biscoveanu:2022qac, Adamcewicz:2023mov, Heinzel:2023hlb, Heinzel:2024hva, Pierra:2024fbl, Franciolini:2022iaa, Antonini:2025zzw, Adamcewicz:2022hce, Berti:2025usa}
that could elucidate the astrophysical formation channels of merging \acp{BBH}.
Also, population inference methods in \ac{GW} astronomy must often contend with systematic uncertainties which grow with the number of observations \cite{Talbot:2023pex, Gaebel:2018poe, jack_math_tome} and require additional computational resources to ameliorate \cite{Wolfe:2025yxu}.
Therefore, population inference grows ever more computationally intensive
from the accumulating complexity of the data.

A natural strategy to understand complex data is to use Bayesian updating
to break the analysis problem down into smaller, simpler parts.
Since the posterior from one set of observations already contains all the relevant information that can be learned from that data (for a given prior), there is in principle no disadvantage to iteratively updating our inference as new observations come in.
As E.~T.~Jaynes reminds us,
``one man's prior probability is another man's posterior probability''\footnote{Equivalently, in this work ``one person's posterior probability is a later person's prior probability.''} \cite{Jaynes:2003jaq}.

However, all current analyses (to our knowledge) adopt relatively uninformed priors on population model parameters
and repeatedly reanalyze the entire catalog of gravitational-wave sources as new events arrive.
Thus, \ac{GW} catalogs may outgrow the available memory on hardware accelerators
(e.g., GPUs)---which have dramatically reduced the computation time required for population
inference---if GPU memory no longer grows exponentially in the coming decades
\cite{doi:10.1126/science.aam9744}.
Updating the population with smaller samples of signals could avoid these computational bottlenecks.
Bayesian updating could also enable efficient updates to \ac{GW} rate estimates and population
models during an \ac{LVK} observing run.
In turn, search pipelines could update the probability that an event is astrophysical
($p_{\rm astro}$; see \citet{Banagiri:2023ztt} and references therein)
or be tuned for sensitivity to particular populations \cite{Dent:2013cva},
analysts could use the latest population-informed priors for computationally expensive analyses of long signals, e.g., binary neutron star mergers,
and online population updates could be made public to guide astrophysical modeling
of the formation channels of \ac{GW} sources
well in advance of public \ac{LVK} data releases.
These applications motivate us to investigate the practicality of performing
iterative Bayesian updating for gravitational-wave population inference.

Bayesian updates require a tractable posterior density that can be rapidly evaluated given observed data,
to serve as the prior when analyzing subsequent observations.
The typical output of existing analyses,
a finite number of samples drawn from the posterior of population model parameters,
is not immediately usable for this purpose.
While one could fit the distribution of
samples with a density estimator to obtain a tractable density,
recent applications of both simulation-based~\cite{Leyde:2023iof} and variational~\cite{Mould:2025dts} inference to hierarchical population inference imply that we can simultaneously acquire a set of posterior samples given current data \textit{and} a reusable
posterior density that can serve as the prior in subsequent \ac{GW} population analyses.
\citet{Leyde:2023iof} trained a conditional density estimator to fit the population distribution given fixed-size catalogs of \ac{BBH} mergers.
Training with a catalog of 100 sources requires a few hours and inference thereafter takes only seconds.
However, training bakes in fixed assumptions like the physics of gravitational waveforms
or the functional form of the astrophysical population.
Further, recovering the posterior for more than 100 events requires combining the output of multiple density estimators simultaneously;
\citet{Leyde:2023iof} found that noticeable errors accumulate in the final posterior when combining more than $\sim 10$ sub-catalogs at a time.

Recently, \citet{Mould:2025dts} used neural variational methods~\cite{2015arXiv150505770J}
to perform population inference in
seconds on current GW catalogs,
enabling rapid recovery of astrophysical distributions
under different population models or given new observations.
Here, we extend this approach and
seek to evaluate whether sequential Bayesian updates of the population distribution
with neural variational methods
is a reliable approach for population inference.
In Sec.~\ref{sec:framework},
we review hierarchical Bayesian inference and updating in \ac{GW} astronomy.
In Sec.~\ref{sec:methods}, we outline a framework for sequential updates via neural \ac{VI}.
Then, in Sec.~\ref{sec:gwtc4},
we evaluate this methodology by repeating the latest \ac{LVK} population analysis \cite{LIGOScientific:2025pvj}
with sequential updates.
In Sec.~\ref{sec:mock}, we extend neural sequential updates
towards future catalogs under both a low-dimensional phenomenological population model
and a flexible high-dimensional model.
Finally, we summarize our findings in Sec.~\ref {sec:conclusions}
and identify additional applications enabled by neural Bayesian updating.

\section{Population inference} \label{sec:framework}

\subsection{Hierarchical likelihood} \label{sec:hierarchical-likelihood}

Neglecting eccentricity,
binary black holes are characterized by 15 source parameters $\theta$,
including their component masses, spin angular momenta, location and orientation,
and arrival time $t$ at the detector, as measured in the frame of the detector.
Let $R(\theta | \Lambda)$ be the differential
rate of gravitational-wave mergers
in the detector-frame.
This rate has a functional form---the population model---parameterized by $\Lambda$,
which must be chosen prior to inference.
Astrophysical modeling typically considers
the differential merger rate density $\mathcal{R}$ in the frame of sources
at redshift $z$,
which is related to the detector-frame merger rate by
\begin{align} \label{eq:src-frame-com-rate}
    \mathcal{R}(\theta'; z) = R(\theta) \left( \frac{1}{1+z} \frac{\dd V_c}{\dd z} \right)^{-1} \, ,
\end{align}
where $V_c$ is comoving volume
and $\theta'$ is all of $\theta$ except $z$
which we separate to make explicit that $\mathcal{R}$ is not a distribution in $z$
although it does evolve with redshift.

We assume that gravitational waves are generated and arrive at the detector network independently.
For data $d$ observed in a network of gravitational-wave detectors, their likelihood is
\begin{align} \label{eq:single-like}
    p(d, \theta | \Lambda) \propto p(d | \theta) R(\theta | \Lambda) \, ,
\end{align}
where $p(d|\theta)$ is the usual single-event GW likelihood function
\cite{whittle, Cornish:2013nma, Thrane:2018qnx}.
The expected number of detections is
\begin{equation} \label{eq:nexp}
    \bar{N}(\Lambda, T) = \int_T \dd t \, \dd \vartheta \, \dd d \, P(\det | d) p(d | \theta) R(\theta | \Lambda) \, ,
\end{equation}
where $T$ is the period of observation,
$\vartheta$ is all of $\theta$ except $t$,
and $P(\det | d)$ is the probability of assigning a detection (``$\det$'') to a particular piece of data;
the integral over $d$ is taken over all possible noise realizations with added signal $h(\theta)$.
Throughout the rest of this work,
we take $P(\det | d)$ as a Heaviside step function,
selecting all data with a detection statistic---such as \ac{SNR} or \ac{FAR}---surpassing
some threshold.
We emphasize that since $\theta$ contains the arrival time of the \ac{GW}
we have not assumed constant detector sensitivity.

Under the assumption that gravitational-wave events are independent, we can model the occurrence of $N$ observations in the presence of selection effects as an inhomogeneous Poisson process; then, the catalog \cat{} of data containing observed signals and the set of all their source parameters \src{} have a joint density\footnote{Note that Eq.~\eqref{eq:pop-like} is not strictly a ``likelihood'' as it depends on latent variables \src{}. }
\begin{align} \label{eq:pop-like}
    \mathcal{L}(\cat{}, \src{} | \Lambda) \propto e^{-\bar{N}(\Lambda, T)} \prod_{i = 1}^N p(d_i | \theta_i) R(\theta_i | \Lambda) \, .
\end{align}
See \cite{Loredo:2001rx, Messenger:2012jy, 2019MNRAS.486.1086M, Vitale:2020aaz} for derivations of the inhomogeneous Poisson process likelihood in astronomical contexts.

\subsection{Sequential updates} \label{sec:updates}

We begin by dividing an observing period of time $T$
into two disjoint periods of time $T_1$ and $T_2$,
over which we observe data $\cat_{(1,2)}$
and make $N_{(1,2)}$ detections with source parameters $\src_{(1,2)}$,
such that $T = T_1 + T_2$, $N = N_1 + N_2$,
$\src = \{ \src_1, \src_2 \}$
and $\cat = \{ \cat_1, \cat_2 \}$.
Note in Eq.~\eqref{eq:nexp} that the domain of integration separates in detector time, which implies that $\bar{N}(\Lambda, T) = \bar{N}(\Lambda, T_1) + \bar{N}(\Lambda, T_2)$.
As the event-likelihoods $p(d_i | \theta_i) R(\theta_i | \Lambda)$ are independent,
their product in Eq.~\eqref{eq:pop-like} separates between the two observing periods.
Therefore, we have that
\begin{align} \label{eq:factor-pop-like}
    \mathcal{L}(\cat, \src | \Lambda) = \mathcal{L}(\cat_1, \src_1 | \Lambda) \mathcal{L}(\cat_2, \src_2 | \Lambda) \, .
\end{align}
By Bayes' theorem,
and for a chosen prior $\pi$,
the joint posterior on all single-event and population parameters is
\begin{align} \label{eq:tot-post}
    \mathcal{P}(\src{}, \Lambda | \cat{}) &= \frac{ \mathcal{L}(\cat{}, \src{} | \Lambda) \pi(\Lambda) }{ \int \dd \Lambda \, \dd \src{} \, \mathcal{L}(\cat{}, \src{} | \Lambda) \pi(\Lambda) } \, .
\end{align}
Similarly, the posterior after just the first observing period is $\mathcal{P}(\src_1, \Lambda | \cat_1) \propto \mathcal{L}(\cat_1, \src_1 | \Lambda) \pi(\Lambda)$.
With Eq.~\eqref{eq:factor-pop-like}, this implies that we can write the posterior given all the data as
\begin{equation}
\begin{aligned}
    &\mathcal{P}(\src, \Lambda | \cat) \\
    &= \frac{ \mathcal{L}(\cat_2, \src_2 | \Lambda) \mathcal{L}(\cat_1, \src_1 | \Lambda) \pi(\Lambda) }{ \int \dd \src_1 \, \dd \src_2 \, \dd \Lambda \, \mathcal{L}(\cat_2, \src_2 | \Lambda) \mathcal{L}(\cat_1, \src_1 | \Lambda) \pi(\Lambda) } \\
    &= \frac{  \mathcal{L}(\cat_2, \src_2 | \Lambda) \mathcal{P}(\src_1, \Lambda | \cat_1) }{ \int \dd \src_1 \, \dd \src_2 \, \dd \Lambda \, \mathcal{L}(\cat_2, \src_2 | \Lambda) \mathcal{P}(\src_1, \Lambda | \cat_1) } \, , \\
\end{aligned}
\end{equation}
that is, the posterior from previous observations takes the place of the prior when analyzing new observations.
Note that from the second to third lines, the constant normalization of $\mathcal{P}(\src_1, \Lambda | \cat_1)$ cancels between the numerator and denominator.

So far,
we have described a Bayesian update of the ``full hierarchical'' posterior,
showing that we can update our knowledge of the astrophysical distribution
\textit{and} source properties of each event as new data are collected.
However, the posterior (Eq.~\ref{eq:tot-post}) marginalized over $\src{}$ is more commonly studied
in \ac{GW} population analyses
(see Refs.~\cite{Mould:2023eca, Mancarella:2025uat} for fully hierarchical approaches),
so in this work we perform sequential updates of the marginal posterior distribution of $\Lambda$.
The likelihood of a single event with data $d_i$ marginalized over its
source parameters $\theta_i$ is
\begin{align} \label{eq:marg-single-event-like}
    p(d_i | \Lambda) \propto \int \dd \theta_i \, p(d_i | \theta_i) R(\theta_i | \Lambda) \, .
\end{align}
We redefine $\mathcal{L}$ as the population likelihood marginalized over source parameters,
\begin{align} \label{eq:marg-pop-like}
    \mathcal{L}(\cat{} | \Lambda) &= \int \dd \Theta \, \mathcal{L}(\cat{}, \Theta | \Lambda) = e^{-\bar{N}(T)} \prod_{i = 1}^N p(d_i | \Lambda) \, .
\end{align}
Since $\pi$ does not depend on $\Theta$,
the posterior on $\Lambda$ marginalized over $\Theta$ is
$\mathcal{P}(\Lambda | \cat{}) \propto \mathcal{L}(\cat{} | \Lambda) \pi(\Lambda)$.
The single-event likelihoods are still independent after marginalizing
over their source parameters,
so we can factorize $\mathcal{L}(\cat | \Lambda)$
into the marginal likelihoods $\mathcal{L}(\cat_1 | \Lambda)$ and $\mathcal{L}(\cat_2 | \Lambda)$
of the data observed in each period.
In turn,
the posterior on $\Lambda$ given the first observing period is
$\mathcal{P}(\Lambda | \cat_1) \propto \mathcal{L}(\cat_1 | \Lambda) \pi(\Lambda)$,
which can be updated after the second observing period to yield
$\mathcal{P}(\Lambda | \cat) \propto \mathcal{L}(\cat_2 | \Lambda) \mathcal{P}(\Lambda | \cat_1)$.

In general,
if we divide our data into $n$ disjoint segments with durations $T_m$ and data $\cat_m$,
$m = 1, \ldots, n$,
the likelihood of the entire dataset is $\mathcal{L}(\cat | \Lambda) = \mathcal{L}(\cat_1, \ldots, \cat_n | \Lambda) = \prod_{m = 1}^n \mathcal{L}(\cat_m | \Lambda)$.
We notate the likelihood of the $m$th segment as $\mathcal{L}_m := \mathcal{L}(\cat_m | \Lambda)$,
and the posterior given all segments up to and including the $m$th as $\mathcal{P}_m := \mathcal{P}(\Lambda | \cat_1, \ldots, \cat_m)$.
The posterior given the first segment depends on the prior $\pi$.
Thereafter, we can define a sequence of updated posteriors
$\mathcal{P}_m \propto \mathcal{L}_m \mathcal{P}_{m - 1}$.

\subsection{Monte Carlo integration} \label{sec:monte-carlo}

In practice, we estimate the single-event likelihoods in Eq.~\eqref{eq:marg-single-event-like}
as Monte Carlo integrals over sets of $N_{\rm PE}$ single-event posterior samples;
similarly, we estimate $\bar{N}$ in Eq.~\eqref{eq:nexp} as a Monte Carlo integral
over $N_{\rm det}$ observable synthetic sources \cite{Tiwari:2017ndi, 2019RNAAS...3...66F}.
The estimate of the log-likelihood, $\ln \hat{\mathcal{L}}$,
has an associated Monte Carlo variance \var{} \cite{Tiwari:2017ndi, 2019RNAAS...3...66F, Essick:2022ojx, Talbot:2023pex, jack_math_tome}.
Note that \var{} depends on the population parameters,
the set of observable synthetic sources for sensitivity estimation,
and the particular catalog of events being analyzed
as these modify the importance weights in the Monte Carlo approximations to Eqs.~\eqref{eq:nexp} and Eq.~\eqref{eq:marg-single-event-like}.
When \var{} is large, $\ln \hat{\mathcal{L}}$ may be high for population distributions
that are otherwise disfavored by the observed data \cite{Talbot:2023pex, jack_math_tome}.
We penalize the likelihood estimator when $\mathcal{V}$ is large by multiplying it with a taper $\mathcal{T}$, yielding a regularized likelihood estimator $\hat{\mathcal{L}} \mathcal{T}$;
 we take $\ln\mathcal{T}(\Lambda) = -100 [\mathcal{V}(\Lambda) - V]^2$ when $\mathcal{V}(\Lambda) > V$ \cite{Heinzel:2024jlc, Mould:2025dts}
for a variance threshold $V$.
When performing sequential updates,
we will either take $\mathcal{V}$ as an approximation to the variance of the
cumulative log-likelihood (Sec.~\ref{sec:updates-o4a})
or the variance of the log-likelihood for a particular update (Sec.~\ref{sec:mock}).
Either choice means that the taper depends on the update;
we define $\mathcal{T}_m$ as the taper applied to the likelihood estimate
$\hat{\mathcal{L}}_m$.

\section{Neural Bayesian updating} \label{sec:methods}

\subsection{Neural variational inference} \label{sec:neural-vi}

As shown in Sec.~\ref{sec:updates}, updating prior belief for future analyses requires the posterior density from past data.
Here, we use \ac{VI} to tractably estimate the posterior density following each update.
Let $\mathcal{Q}(\Lambda; {\varphi})$ define a family of distributions parameterized by $\varphi$.
Inference proceeds by finding $\varphi$ that minimizes some distance between $\mathcal{Q}$ and a target distribution $\mathcal{P}$.
We take the ``reverse'' \ac{KL} divergence~\cite{kl} as our measure of difference between distributions, where
\begin{align} \label{eq:kl}
    \mathrm{KL}(\mathcal{Q}, \mathcal{P}) = \int \dd \Lambda \, \mathcal{Q}(\Lambda; \varphi) \ln \frac{ \mathcal{Q}(\Lambda; \varphi) }{ \mathcal{P}(\Lambda) } \, .
\end{align}
In this work, $\Lambda$ will take a dimension of at least $10$ and up to $\sim10^3$.
For computational expediency, we evaluate Eq.~\eqref{eq:kl} by Monte Carlo integration with $B$ samples from $\mathcal{Q}$, which leads to a target loss function
\begin{align} \label{eq:loss}
    L(\varphi; \mathcal{{P}}) = \frac{1}{B} \sum_{s = 1}^{B} \ln \frac{ \mathcal{Q}(\Lambda_s; \varphi) }{ \mathcal{P}(\Lambda_s) } \, .
\end{align}
Throughout this work, we use stochastic gradient descent via the \textsc{Adam} optimizer \cite{kingma2014adam}
with a cosine-decay learning rate from a positive initial learning rate $\eta$ to zero to minimize the loss \cite{2017arXiv170807120S}.
We parameterize a variational family $\mathcal{Q}$ with \textit{normalizing flows}: learnable bijections between a simple distribution (in this work, a multivariate uncorrelated Gaussian)
and a more complex target distribution.
In particular, we use block neural autoregressive flows \cite{2019arXiv190404676D},
which parameterize the bijection as a neural network.
The architecture of the block neural autoregressive flow as implemented in
\texttt{flowjax} \cite{ward2023flowjax} is specified by the neural-network depth and block dimension (neural-network width is block dimension times the dimension of the target space) of each bijection,
and the number of bijections (i.e., flow layers).
Throughout this work, we use a neural network depth of one.
We also found more than one flow layer to be numerically unstable during sequential updates.
We will tune the block dimension to increase the neural network width and, in turn, increase the expressiveness of the flow.

\subsection{Neural sequential updates} \label{sec:neural-sequential-updates}

We perform Bayesian updates with neural \ac{VI} in the following manner:
first, we train $\mathcal{Q}(\Lambda; \varphi)$ to approximate
the posterior given some
initial catalog of observations,
$\mathcal{P}_1 \propto \hat{\mathcal{L}}_1 \pi \mathcal{T}_1$.
When we fit $\mathcal{Q}$ to $\mathcal{P}_1$, we call its parameters $\varphi_1$
and define $\mathcal{Q}_1 = \mathcal{Q}(\Lambda; \varphi = \varphi_1)$.
According to Section~\ref{sec:updates}, $\mathcal{Q}_1$ becomes the prior when analyzing a second segment of data,
approximating the cumulative posterior estimate 
$\mathcal{P}_2 \propto \hat{\mathcal{L}}_1 \hat{\mathcal{L}}_2 \mathcal{T}_1 \mathcal{T}_2 \pi$
as
$\hat{\mathcal{L}}_2 \mathcal{Q}_1 \mathcal{T}_2$.
To this latter density, we fit another variational approximant
with parameters $\varphi_2$ which we call $\mathcal{Q}_2$;
when training $\varphi_2$, we initialize it at $\varphi_1$
to reuse information from the previous update.
This makes inference more efficient as (1) previous observations are not reanalyzed after each update and (2) previous posteriors are already closer to the next posterior update than either the original prior or a random initialization.
In this manner,
we recursively update our posteriors for $m > 1$ by training $\mathcal{Q}_m$
to approximate $\hat{\mathcal{L}}_m \mathcal{T}_m \mathcal{Q}_{m - 1}$.
We summarize our notation in Tab.~\ref{tab:summary}.

\begin{table}[h]
    \centering
    \begin{tabular}{p{2cm} p{1.7cm} p{2.2cm} p{2cm}}
        \toprule
        \toprule
        Variational \newline approximant & Training \newline density & Regularized \newline estimator & Cumulative \newline posterior \\
        \midrule
        $\mathcal{Q}_1$ & $\hat{\mathcal{L}}_1 \pi \mathcal{T}_1$ & $\hat{\mathcal{L}}_1 \pi \mathcal{T}_1$ & $\mathcal{L}_1 \pi$ \\
        $\mathcal{Q}_2$ & $\hat{\mathcal{L}}_2 \mathcal{Q}_1 \mathcal{T}_2$ & $\hat{\mathcal{L}}_1 \hat{\mathcal{L}}_2 \mathcal{T}_1 \mathcal{T}_2 \pi$ & $\mathcal{L}_1 \mathcal{L}_2 \pi$ \\
        \vdots & \vdots & \vdots & \vdots \\
        $\mathcal{Q}_n$ & $\hat{\mathcal{L}}_n \mathcal{Q}_{n - 1} \mathcal{T}_n$ & $\pi \prod_{m = 1}^n \hat{\mathcal{L}}_m \mathcal{T}_m$ & $\pi \prod_{m = 1}^n \mathcal{L}_m$ \\
        \bottomrule
        \bottomrule
    \end{tabular}
    \caption{
    From left to right, the variational approximant is fit to the training density
    which approximates a regularized Monte Carlo estimate of the (unnormalized)
    posterior density.
    }
    \label{tab:summary}
\end{table}

Since we approximate the updated prior for updates $m > 1$ it is possible to accumulate error over many successive updates.
Therefore, after each update, we evaluate the importance sampling weights of $\mathcal{Q}_m$ relative to two distributions:
\begin{itemize}
    \item The training density; the weights $\hat{\mathcal{L}}_m \mathcal{Q}_{m - 1} \mathcal{T}_m / \mathcal{Q}_{m}$ are used to evaluate convergence of the training procedure.
    \item The regularized cumulative posterior estimator; the weights $\mathcal{P}_{m} / \mathcal{Q}_m$ (at samples below the variance threshold) are used to evaluate accuracy over many updates.
\end{itemize}
Both sets of weights are computed over the same set of $10^4$ samples from $\mathcal{Q}_m$.
From each set of weights, we compute two convergence criteria:
the Pareto shape parameter $\hat{k}$ of the weight distribution \cite{2015arXiv150702646V}
and the effective sampling efficiency $\varepsilon$ \cite{kish1995}.
We will take $\hat{k} \lesssim 0.7$ and $\varepsilon \gtrsim 1$\%
as indicative of convergence,
in the sense that the variational approximant is an efficient proposal distribution
for importance sampling to the desired posterior
(note that this threshold on $\hat{k}$ is a heuristic one identified by \citet{2015arXiv150702646V}).
For each experiment in this work,
we also show draws from the variational approximant after the final, $n$th update which we have been importance sampled according to Pareto-smoothed weights relative to the cumulative posterior;
when $\hat{k} \lesssim 0.7$, this procedure corrects discrepancies between $\mathcal{Q}_n$ and $\mathcal{P}_n$ (see \citet{Mould:2025dts} for further discussion).

Stochastic gradient descent requires an (automatically) differentiable target density;
when performing updates, that requires $\mathcal{Q}$ be differentiable with respect to $\Lambda$.
While this is theoretically true, the block neural autoregressive bijection is not analytically invertible; instead, $\mathcal{Q}(\Lambda)$ is computed by numerically inverting the bijection via a greedy bisection search \cite{ward2023flowjax}, which is not automatically differentiable out of the box.
We fork \texttt{flowjax}\footnote{ \url{https://github.com/noahewolfe/flowjax}.}, defining the derivative through the numerical inverse by the implicit function theorem \cite{kidger2021equinox}.
We use \texttt{lineax} \cite{lineax2023} for memory-efficient computation of the Jacobian-vector product.
Our fork also reimplements
construction
of the weight matrices with Kronecker products,
reducing the time to instantiate large block-neural autoregressive flows by a few orders of magnitude.

\section{Real catalogs} \label{sec:gwtc4}

We first demonstrate neural sequential updates
with the most recent \ac{LVK} catalog of \ac{BBH} mergers.
We begin from a posterior on population model parameters given the next-to-latest catalog,
the third gravitational-wave transient catalog (GWTC-3)\acused{GWTC} \cite{KAGRA:2021vkt},
as our prior when analyzing the \ac{BBH} observed in the first part of the \acp{LVK} fourth observing run, O4a \cite{LIGOScientific:2025slb}.
We perform sequential updates at three cadences:
with all of the events observed during O4a,
updating month-by-month,
and as each event arrives.

\subsection{Source selection and sensitivity estimation}

Evaluating the Poisson likelihood in the presence of selection effects
requires us to estimate the number $\bar{N}$ of sources we expect to
observe with a detection statistic above some threshold.
We estimate $\bar{N}$ by Monte Carlo integration over synthetic signals
added to real detector data as released by the
\ac{LVK} \cite{Essick:2025zed, o4a-sensitivity-zenodo, gwtc4-cumulative-sensitivity-zenodo}.
Following \citet{LIGOScientific:2025pvj},
we select sources observed with $\mathrm{SNR} > 10$ during the \acp{LVK}
first and second observing runs (O1 and O2),
and sources with $\mathrm{FAR} < 1\,\rm yr^{-1}$ thereafter.

We perform sequential updates of the population,
selecting \ac{BBH} observed during O4a at three cadences:
from the beginning to the end of O4a;
from the start and end time of each calendar month
(we combine eight days at the end of May 2023 with June 2023 as no \ac{BBH} were observed during May 2023);
and after each event arrived during O4a.
To calculate $\bar{N}$ during a particular observing period,
we downselect only the detectable, synthetic signals generated
in that same period.
We always exclude the engineering run prior to O4a when estimating $\bar{N}$.
When performing updates as each event arrived,
we select the synthetic signals generated since the last event;
for example, when updating the population after the arrival of GW231123\_135430 (hereafter GW231123) \cite{LIGOScientific:2025rsn},
we use the synthetic signals since GW231119\_075248.
We directly convert event labels into GPS times for event selection
with \texttt{gwpy} \cite{gwpy}.
We also compute the cumulative posterior density given all the data observed
so far, for which we estimate $\bar{N}$ using synthetic signals from the beginning of O1
through the latest update.

The public synthetic signals have been generated
in distinct sets for O1, O2, O3, and each month of O4a.
The duty cycle of the detector is not the same in each of these periods;
therefore, whenever we select synthetic signals from multiple sets
we reweight them according to a uniform prior in detector-frame time
following Eq.~(20) in \citet{2021RNAAS...5..220E}.
This requires the total number of synthetic signals (detected and non-detected)
generated in each set,
which we record in App.~\ref{app:ninj}.

\subsection{Fitting \ac{GWTC}-3} \label{sec:gwtc3}

We first analyze the 69 \ac{BBH} observed through the \acp{LVK} third observing run (O3).
Starting from Eq.~\eqref{eq:src-frame-com-rate},
we model the source-frame, differential merger rate density as
\begin{equation} \label{eq:model-explicit-rate}
    \mathcal{R}(\theta'; z) = K \left( \frac{1}{1 + z} \frac{\dd V_c}{\dd z} \right)^{-1} p(\vartheta | \Lambda) \, ,
\end{equation}
where $K$ is the rate of \ac{BBH} mergers (observed and unobserved)
in the frame of the detector \cite{Essick:2023upv},
and $\vartheta$ is all of $\theta$ except $t$.
For the shape $p(\vartheta | \Lambda)$ of the population,
we adopt a lightly modified version of a strong \ac{BBH} population model from \citet{LIGOScientific:2025pvj},
which assumes a tapered and broken power law with two Gaussian peaks in $m_1$,
a tapered power law in $q$,
a power law in $1 + z$,
independent and identical truncated Gaussians in $\chi_{1,2}$,
and a mixture between an isotropic distribution and a Gaussian peak in $\cos \tau_{1,2}$
(note that the peak location is \textit{not} fixed at $\cos \tau_{1,2} = 1$).
For simplicity, we adopt the same minimum mass $m_{\min{}}$ and low-mass smoothing scale $\delta_m$
for primary and secondary masses,
as opposed to \citet{LIGOScientific:2025pvj}
which allowed $m_{\min{}}$ and $\delta_m$ to differ between component masses.
We also truncate the $m_1$ distribution at $300\,M_\odot$, as in \citet{LIGOScientific:2025pvj}.
This model has a total of $D = 19$ parameters.
Priors can be found in Tab.~\ref{tab:gwtc-priors} of Appendix~\ref{app:priors}.
We adopt a variance threshold of $V = 1$.

For \ac{VI},
we use a flow with a block dimension of 8.
We initialize the flow to the prior,
by training on prior draws for $10^3$ steps with a batch size of $B = 10^4$,
starting from a learning rate of $\eta = 1$ and decaying to zero
without gradient clipping.
Then,
from that initialization we train the approximant $\mathcal{Q}_\mathrm{O3}$
for $10^4$ steps with a batch size of $B = 10$ starting from a learning rate of
$\eta = 10^{-1}$ and decaying to zero,
clipping gradients to a maximum norm of 1.
Training takes 3 minutes.
Our variational approximant yields $\hat{k} \approx 0.66$ and $\varepsilon = 10.5\%$,
indicating convergence.

\subsection{Updating through O4a} \label{sec:updates-o4a}

\begin{figure}
    \centering
    \includegraphics[width=0.99\linewidth]{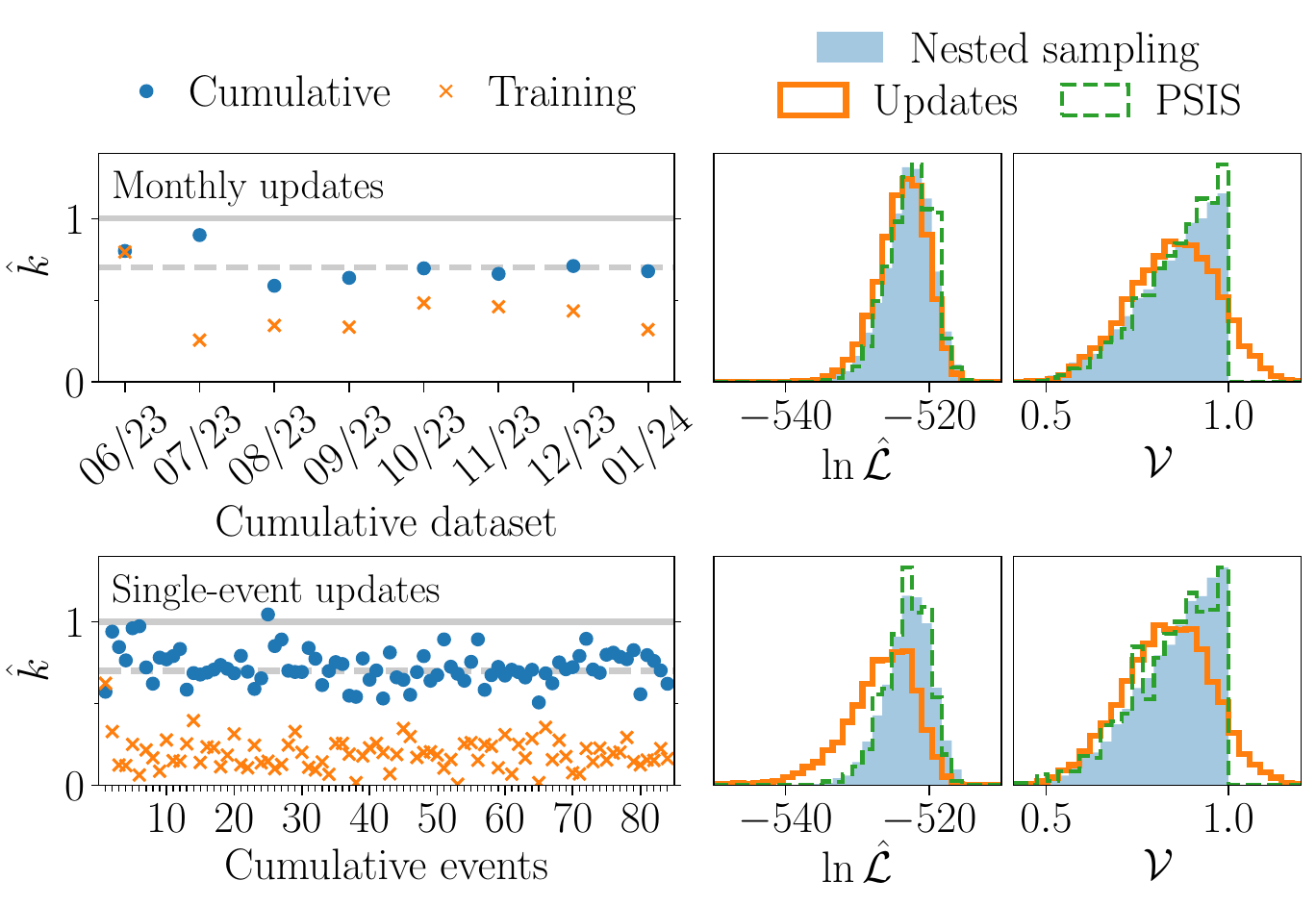}
    \caption{
    Convergence of sequential updates starting from \ac{GWTC}-3 and adding \ac{BBH} observed in O4a.
    We show the Pareto smoothing shape parameter $\hat{k}$ as a function of update (left)
    when we update month-by-month (top) or as each event arrives (bottom).
    We compute $\hat{k}$ relative to the cumulative posterior (circles)
    and the target density during training (crosses).
    We include a dashed gray line at $\hat{k} = 0.7$, a reference threshold for convergence \cite{2015arXiv150702646V}.
    In the right two columns, we compare posterior distributions
    on log-likelihood estimator $\ln \hat{\mathcal{L}}$ and variance $\mathcal{V}$
    drawn according to nested sampling (blue, filled),
    the variational approximant after the final update (orange),
    and Pareto-smoothed importance sampling (PSIS) of the variational approximant (green).
    }
    \label{fig:gwtc4-convergence}
\end{figure}

\begin{figure*}
    \centering
    \includegraphics[width=0.99\linewidth]{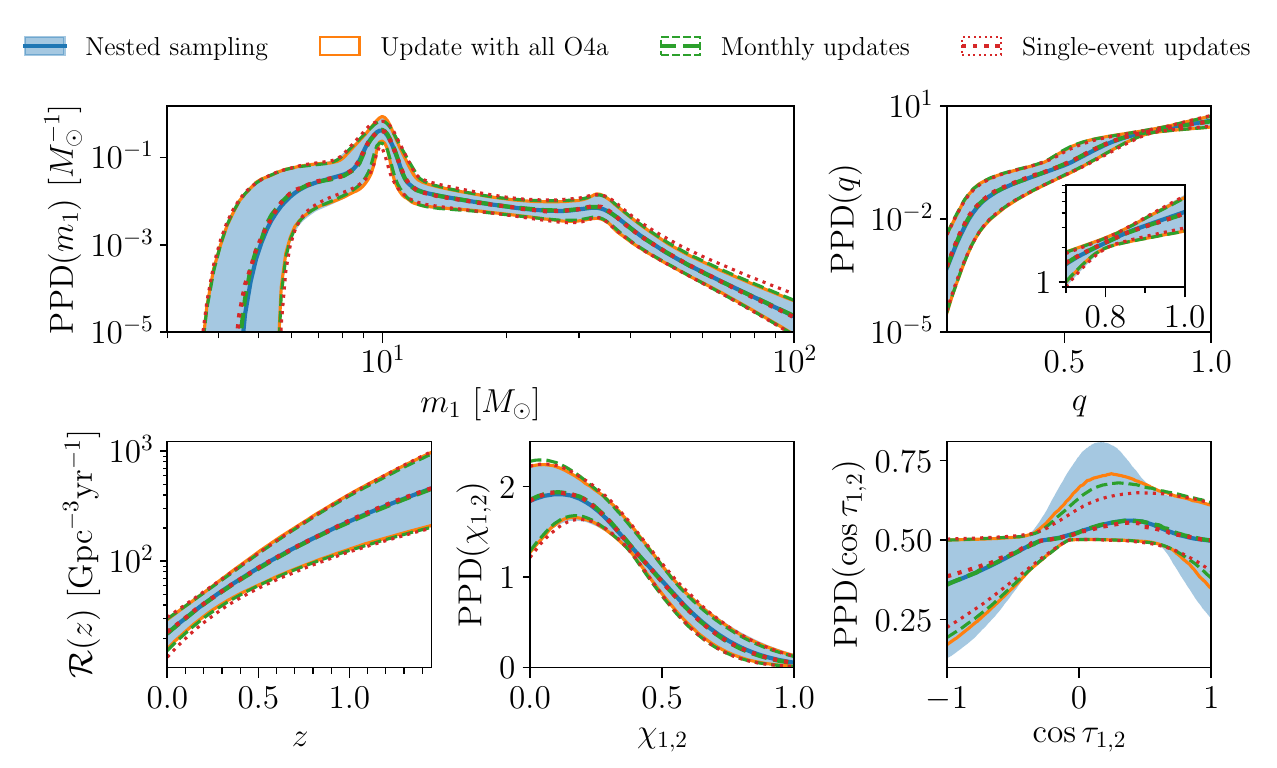}
    \caption{
    \acsp{PPD} in primary mass (top left), mass ratio (top right), spin magnitude (bottom center), and tilt (bottom right) given all 153 \acp{BBH} in \ac{GWTC}-4.
    We also show the posterior on the source-frame merger rate density over redshift (bottom left).    
    We compare results obtained with nested sampling (blue, filled)
    and sequentially updating with all of O4a (orange),
    month by month (green),
    and event by event (red).
    Thin lines or shading enclose the 90\% credible region and thick lines denote medians;
    we exclude the median of the update with all of O4a
    as it was nearly identical to the median reported by nested sampling.
    In the top right, we include an inset showing $\mathrm{PPD}(q)$ at $q > 0.7$.
    }
    \label{fig:gwtc-updates-all}
\end{figure*}

We now proceed to update this posterior
with the 84 \ac{BBH} observed in O4a with $\mathrm{FAR} > 1\,\rm yr^{-1}$ \cite{LIGOScientific:2025slb}.
To generate results without \ac{VI} to compare with,
we analyze the \ac{BBH} through \ac{GWTC}-4
using nested sampling with \texttt{dynesty} \cite{Speagle:2019ivv, sergey_koposov_2025_17268284}
via \texttt{bilby} \cite{bilby_paper};
we use 1000 live points,
the \texttt{acceptance-walk} method to sample within likelihood contours
over a fixed number \texttt{naccept} of 5 autocorrelation times.
We use the same priors and variance threshold as our analysis of \ac{GWTC}-3.

First, we perform the simplest update---taking $\mathcal{Q}_\mathrm{O3}$ as our prior
when analyzing all of the \ac{BBH} observed in O4a at once.
We use the same flow architecture and training settings as in Sec.~\ref{sec:gwtc3},
initializing the variational approximant $\mathcal{Q}_{\mathrm{O4a}}$
to $\mathcal{Q}_{\mathrm{O3}}$.
We adopt a taper $\mathcal{T}$ on the Monte Carlo variance of the log-likelihood
of only the O4a events with a threshold of $V = 1$.
After training, we compute importance weights of $\mathcal{Q}_\mathrm{O4a}$ relative to the cumulative population posterior,
that is, given all 153 events in \ac{GWTC}-4,
and we find $\hat{k} = 0.62$ and $\varepsilon = 3.7$\%.

We then perform $n = 7$ sequential updates month-by-month,
and then $n = 84$ updates following the individual events in the order that they 
were observed.
We make two modifications to the learning task when performing these updates.
First, we taper on the variance $\mathcal{V}$
of the cumulative log-likelihood for each update,
as opposed to the variance of the log-likelihood of only the new data
in the update,
as we found this improved the number of samples with $\mathcal{V} < 1$.
To avoid expensive computation of the cumulative variance
(which requires effectively the same calculations as the log-likelihood estimator)
we efficiently bound $\mathcal{V}$ by decomposing the variance into contributions from each update
and fitting the cumulative variance from previous updates with a neural network;
see App.~\ref{app:var-approx} for additional details.

Second, we coarsely tune the flow size and learning schedule,
repeating each sequence of updates over a grid of flow size
and learning schedule,
spanning block dimensions of $\{ 2, 4, 8 \}$
and initial learning rates $\eta$ of $\{ 10^{-1}, 10^{-2}, 10^{-3} \}$
decaying to zero, along with a constant learning rate of $10^{-3}$.
Otherwise, we adopt the same training settings as in Sec.~\ref{sec:gwtc3}.
Whenever we use a smaller flow than for $\mathcal{Q}_\mathrm{O3}$,
we initialize the flow before the first O4a update
by fitting it to $\mathcal{Q}_\mathrm{O3}$.
Going forward,
we show the results which achieved the lowest $\hat{k}$ relative to
the cumulative posterior after the final update.
For month-by-month updating,
the best final results were obtained with a block dimension of 8 and $\eta = 10^{-2}$,
achieving $\hat{k} = 0.68$ and $\varepsilon = 4.2\%$;
for single-event updating,
the best results were obtained with a block dimension of 8 and $\eta = 10^{-3}$
achieving $\hat{k} = 0.62$ and $\varepsilon = 3.0\%$.

In Fig.~\ref{fig:gwtc4-convergence},
we show the convergence of sequential month-by-month (top)
and single-event updates (bottom).
When we evaluate importance weights relative to the training density
for the $m$th update
($\mathcal{L}_m \mathcal{T}_m \mathcal{Q}_{m-1}$; cf. Sec.~\ref{sec:neural-sequential-updates}),
we find $\hat{k} \lesssim 0.7$ after nearly every update,
indicating that the training has converged to the posterior under the previous approximate posteriors after each update.
We also find that $\hat{k}$ relative to the cumulative posterior is always
larger than relative to the training density;
this may follow from our initialization procedure,
as we initialize $\mathcal{Q}_m$ to the data-informed prior $\mathcal{Q}_{m - 1}$
which may dominate the training density for some updates.
Over all monthly updates, we see that $\hat{k} < 1$ and near or below 0.7;
similarly, over most single-event updates $\hat{k} < 1$ and is often near or
below 0.7.

While we do not satisfy the criterion proposed by \citet{2015arXiv150702646V}
at every update,
our variational approximation to the posterior
remains accurate enough to reproduce inference with all of \ac{GWTC}-4
after importance sampling.
In the right columns of Fig.~\ref{fig:gwtc4-convergence} we
inspect the distributions of log-likelihood estimates $\ln \hat{\mathcal{L}}$
and variance $\mathcal{V}$ over posterior samples.
After seven monthly updates,
our variational approximant agrees with nested sampling
except in the tails of the $\ln \hat{\mathcal{L}}$ distribution.
It does not perfectly capture the hard threshold on $\mathcal{V}$ used in nested sampling,
although this same behavior was seen with \ac{VI} in \citet{Mould:2025dts} (cf. their Fig. 2);
we apply a hard variance cutoff when importance sampling
and recover the same distribution of $\ln \hat{\mathcal{L}}$ and $\mathcal{V}$
as nested sampling, with minor sampling noise.
Performing single-event updates is more unstable;
by the 84th update,
the distribution of $\ln \hat{\mathcal{L}}$ under the variational approximant
has converged to a similar location but with a heavier tail towards smaller likelihoods.
After importance sampling, we recover the same $\ln \hat{\mathcal{L}}$ and $\mathcal{V}$
distributions as nested sampling albeit with noticeable sampling noise.

We summarize our results in Fig.~\ref{fig:gwtc-updates-all},
where we compare 90\% credible intervals and medians of the marginal \acp{PPD}
in $m_1$, $q$, $\chi_{1,2}$ and $\cos \tau_{1,2}$
returned by nested sampling
and sequential updates at the three cadences (with all of O4a, monthly, and as each event arrives).
We also show the posterior on the source-frame, differential merger rate density $\mathcal{R}$
at each redshift $z$.
All posteriors have been reweighted with \ac{PSIS} after excluding samples with $\mathcal{V} \geq 1$.
Updating with all of O4a or month-by-month return nearly identical
median and 90\% credible regions for the population distributions of $m_1$, $q$, and $z$
as returned by nested sampling.
Updating as each event arrives,
we see slightly more noticeable discrepancies relative to nested sampling with the entire catalog;
for example, the final single-event update slightly underestimates
the 5th percentile in $\mathcal{R}(z)$, especially at $z < 0.5$.
To visually resolve the peaks at $\sim 10\,M_\odot$ and $\sim 35\,M_\odot$
we only show the $m_1$ distributions up to $100\,M_\odot$;
however, the importance-sampled posteriors returned by sequential updates
agree with nested sampling out to the maximum of $300\,M_\odot$.
Event-by-event updating slightly overestimates the 95th percentile of the $m_1$
distribution beyond $\sim 50\,M_\odot$.
These discrepancies are still relatively minor however.
Turning to spin magnitudes,
updating with all of O4a matches the median and 90\% credible region returned by nested sampling;
similarly, monthly and single-event updates nearly match the nested sampling result,
although the final monthly update slightly overestimates the 5th and 95th percentiles at $\chi_{1,2} \lesssim 0.2$.
Most strikingly,
all updates fail to reproduce the posterior on spin tilt distributions,
in particular underestimating the 95th percentile near $\cos \tau_{1,2} \sim 0$
and overestimating the 5th percentile near $\cos \tau_{1,2} \sim 1$.
Misestimation of the location and width of the 90\% credible interval
becomes worse as the updating cadence increases,
particularly when performing single-event updates.

Previously,
\citet{Vitale:2025lms} noted that the recovered $\cos \tau_{1,2}$ distribution,
particularly the peak location,
is subject to noticeable Poisson variability under different \ac{GW} catalog realizations
owing to typically poor measurements of \ac{BBH} spin tilts (see similarly \citet{Miller:2024sui} and \citet[App.~B]{Okounkova:2021xjv}).
Recently,
\citet{Wolfe:2025yxu} noted that only a small fraction of events with the highest \acp{SNR} contribute to population-level measurements of spins.
Taken together, 
these may hint that single-event sequential updates of the spin distribution
are dominated by variance during the training procedure;
in particular,
variance of the Monte Carlo estimate of the loss in Eq.~\eqref{eq:loss} and its gradient.
In turn,
the variational approximant may consistently lie on the edge of convergence---or appear unconverged---at most updates
(cf. Fig.~\ref{fig:gwtc4-convergence}) because the spin-tilt distribution
is challenging to consistently recover.
Further, these results add to previous observations that the recovered
spin-tilt distribution is strongly impacted by Monte Carlo variance (e.g., \citet{Golomb:2022bon});
as such, strong conclusions about the astrophysical population relying on spin tilts should be viewed with caution.
It may be possible to ameliorate these issues by only updating the catalog with data that are expected to be more informative, e.g., events with high \ac{SNR} \cite{Wolfe:2025yxu}.

\subsection{Identifying impactful events}

\begin{figure}
    \centering
    \includegraphics[width=0.8\linewidth]{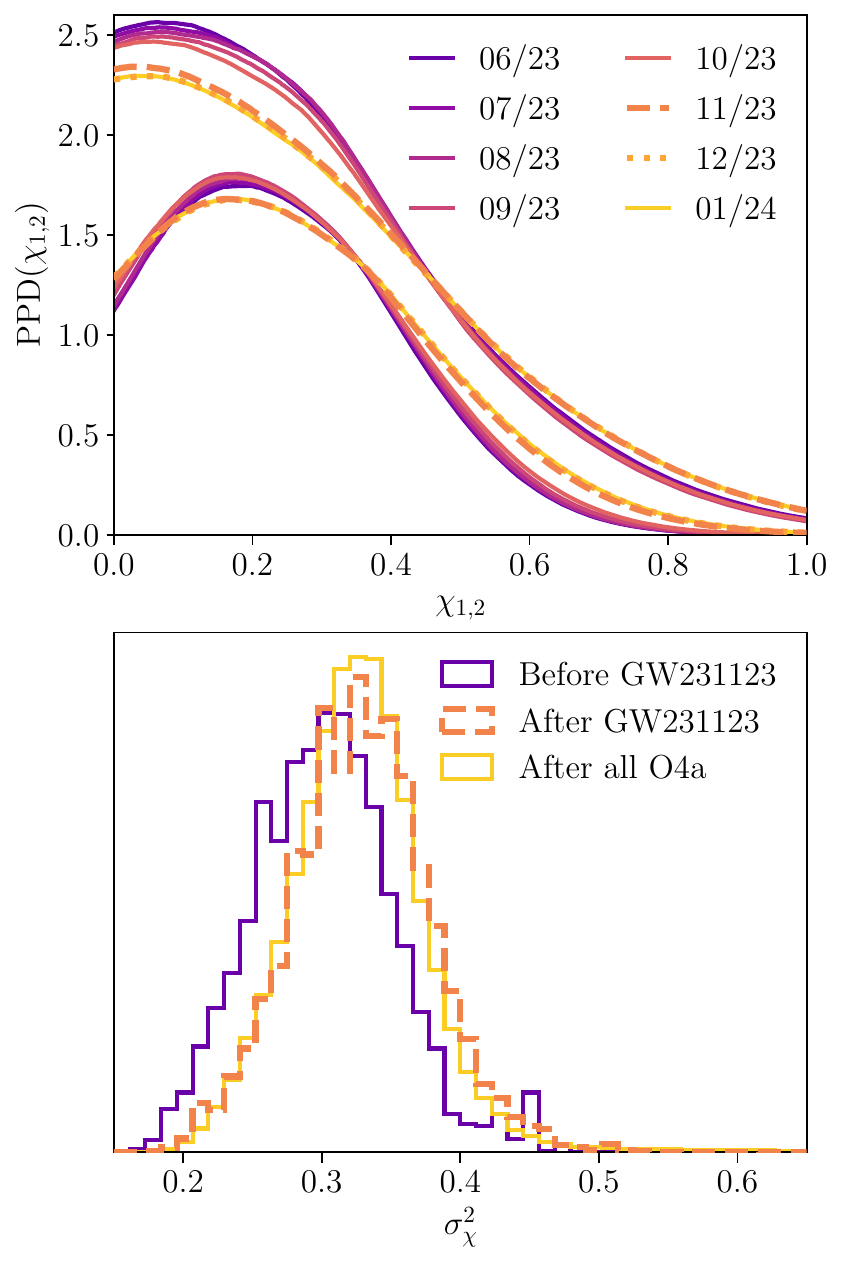}
    \caption{
    \acsp{PPD} in spin magnitude after each month of O4a (top)
    after Pareto-smoothed importance sampling;
    we denote later months with warmer colors and emphasize
    November 2023 and December 2023 with dashed and dotted lines, respectively.
    Marginal importance-sampled posteriors on the width $\sigma_\chi^2$
    of the spin magnitude distribution (bottom) are shown
    before observing GW231123 (purple),
    immediately after GW231123 (orange, dashed),
    and after all of O4a (yellow).
    }
    \label{fig:gw231123}
\end{figure}

Sequential updating of population inference naturally allows us to
identify events which strongly guide our understanding of the \ac{GW} population.
We emphasize that such statements are conditioned on all previous observations.
Here, we briefly outline this use case of sequential updating
by inspecting the spin distribution before and after the observation of GW231123,
a \ac{BBH} merger with strong support for extremal spins \cite{LIGOScientific:2025rsn}.
In the top panel of Fig.~\ref{fig:gw231123},
we plot \acp{PPD} in $\chi_{1,2}$ after each month of O4a;
the events observed in November 2023 reduce support at $\chi_{1,2} \lesssim 0.5$
while increasing it at $\chi_{1,2} \gtrsim 0.5$;
our measurement of the population also grows more certain at larger spins,
with the 90\% credible region shrinking relative to October 2023 at all $\chi_{1,2} \gtrsim 0.7$
(similarly to Fig.~7 in \citet{LIGOScientific:2025pvj} when comparing analyses of \ac{GWTC}-3 and \ac{GWTC}-4).
Inspecting population inference as each event arrives,
we find that GW231123 appears to drive our inference towards
broader spin magnitude distributions.
In the bottom panel of Fig.~\ref{fig:gw231123}, we show the marginal posterior on the width $\sigma_\chi^2$ of the
spin magnitude distribution before observing GW231123, immediately after
that observation, and at the end of O4a
(note that we draw $10^5$ samples,
over $10^4$ used elsewhere,
to clearly resolve each of these histograms).
Before observing GW231123, we infer $\sigma_\chi^2 = 0.30^{+0.08}_{-0.08}$,
(marginal median and 90\% credible interval after \ac{PSIS}),
whereas after we infer $\sigma_\chi^2 = 0.33^{+0.08}_{-0.08}$.
After all of O4a, we infer $\sigma_\chi^2 = 0.33^{+0.12}_{-0.08}$.
Therefore, while GW231123 may guide our inference towards broader
spin distributions,
subsequent observations of merging \ac{BBH} also support similar hypotheses.
Deducing whether GW231123 \textit{alone} strongly determines the spin-magnitude distribution would require inspection of posterior correlations between the population parameters and the source properties of individual events \cite{Moore:2021xhn, Mancarella:2025uat}.

\section{Future catalogs} \label{sec:mock}

\subsection{Mock catalog}

In Sec.~\ref{sec:updates-o4a},
we found that neural sequential updates successfully reproduced
marginal distributions in \ac{BBH} masses, redshift, and spin magnitude
achieved with nested sampling given all of \ac{GWTC}-4 at once.
However,
we found that sequential updates struggled to reproduce the spin-tilt
distribution
and that these errors grew worse when updating with fewer events at a time,
possibly related to poor measurements of the spin tilts of most
observed \ac{BBH}.
Here,
we perform sequential updates with a mock catalog while adopting
a high \ac{SNR} threshold,
to test whether updating with more informative events ameliorates
errors in the tilt distribution
and to evaluate the performance of neural sequential updates
in future observing runs.

We use the mock catalog of \ac{BBH} parameter estimation and sensitivity estimates
from \citet{Wolfe:2025yxu}.
Simulated sources are drawn according to a phenomenological population model
from \citet{KAGRA:2021duu}.
This model assumes: a power law in \ac{BH} primary masses $m_1$ tapered at low masses and with an added Gaussian peak;
a power law in mass ratios $q$ tapered at low masses;
component spin magnitudes $\chi_{1,2}$ following nonsingular beta distributions;
a mixture between isotropic and aligned Gaussian distributions of component spin tilts $\cos \tau_{1,2}$;
and a source-frame merger rate that evolves as a power law in $1 + z$ over redshift $z$.
The exact population parameters for the simulated catalog can be found in \citet[App.~A]{Wolfe:2025yxu}.
This mock catalog has 1600 sources with $\mathrm{SNR} > 11$,
roughly comparable to the expected number of observed \ac{BBH} by the end of the \acp{LVK}
next observing run, O5 \cite{Kiendrebeogo:2023hzf}.
In the following experiments, we use a limited high-\ac{SNR} subset of $N_{\rm obs} = 54$ mock sources
with $\mathrm{SNR} > 30$.
In all experiments
we adopt a variance threshold of $V = 4$ following \citet{Wolfe:2025yxu}
and we penalize variance of the update log-likelihood
(as opposed to the cumulative likelihood as in Sec.~\ref{sec:updates-o4a}).
Finally, we assume a total observing time of $T = 10$\,yr,
implying a local merger rate density of $\sim 15.4\,\rm Gpc^{-3} \, yr^{-1}$ \cite{Wolfe:2025yxu, Mould:2025dts}.
Since the mock catalog assumes a constant detector sensitivity,
we randomly assign a detection time within 10\,yr to each event.

\subsection{Strongly-modeled population} \label{sec:strong}

We begin with a strongly-modeled approach,
explicitly modeling the source-frame differential merger rate density
(cf. Eq.~\ref{eq:model-explicit-rate})
and adopting the same population model
for inference that our simulated sources are drawn from (although we include an additional taper at high masses).
Under this model, $\Lambda$ has dimension $D = 15$.
Priors on $\Lambda$ are shown in Tab.~\ref{tab:strong-priors}.
For comparison, we analyze all $N_{\rm obs} = 54$ simulated sources with nested sampling
using the same settings as in Sec.~\ref{sec:updates-o4a}.

We repeat analysis of the same data via sequential updates,
starting from the prior.
For each update, the initial learning rate is $\eta = 10^{-1}$;
we use $10^4$ training steps;
a batch size of $B = 10$;
and we clip gradients to a maximum norm of 1.
We perform $n = 54, 18, 6$ and 2 sequential updates
with subcatalogs of 1, 3, 9, and 27 events.
When performing 54 single-event updates with $\eta = 10^{-1}$,
the variational approximant after the final update achieves
$\hat{k} \sim 0.74$ and $\varepsilon \sim 0.75 \%$
relative to the cumulative posterior.
To improve convergence,
we tune the training parameters of \ac{VI}
over the coarse grid of flow size and learning schedule in Sec.~\ref{sec:updates-o4a}.
Maintaing a block size of 8 but reducing the inital learning rate to $\eta = 10^{-2}$
yields the smallest cumulative $\hat{k}$ after all single-event updates,
with $\hat{k} \sim 0.59$ and $\varepsilon \sim 0.83\%$.

\begin{figure}
    \centering
    \includegraphics[width=0.99\linewidth]{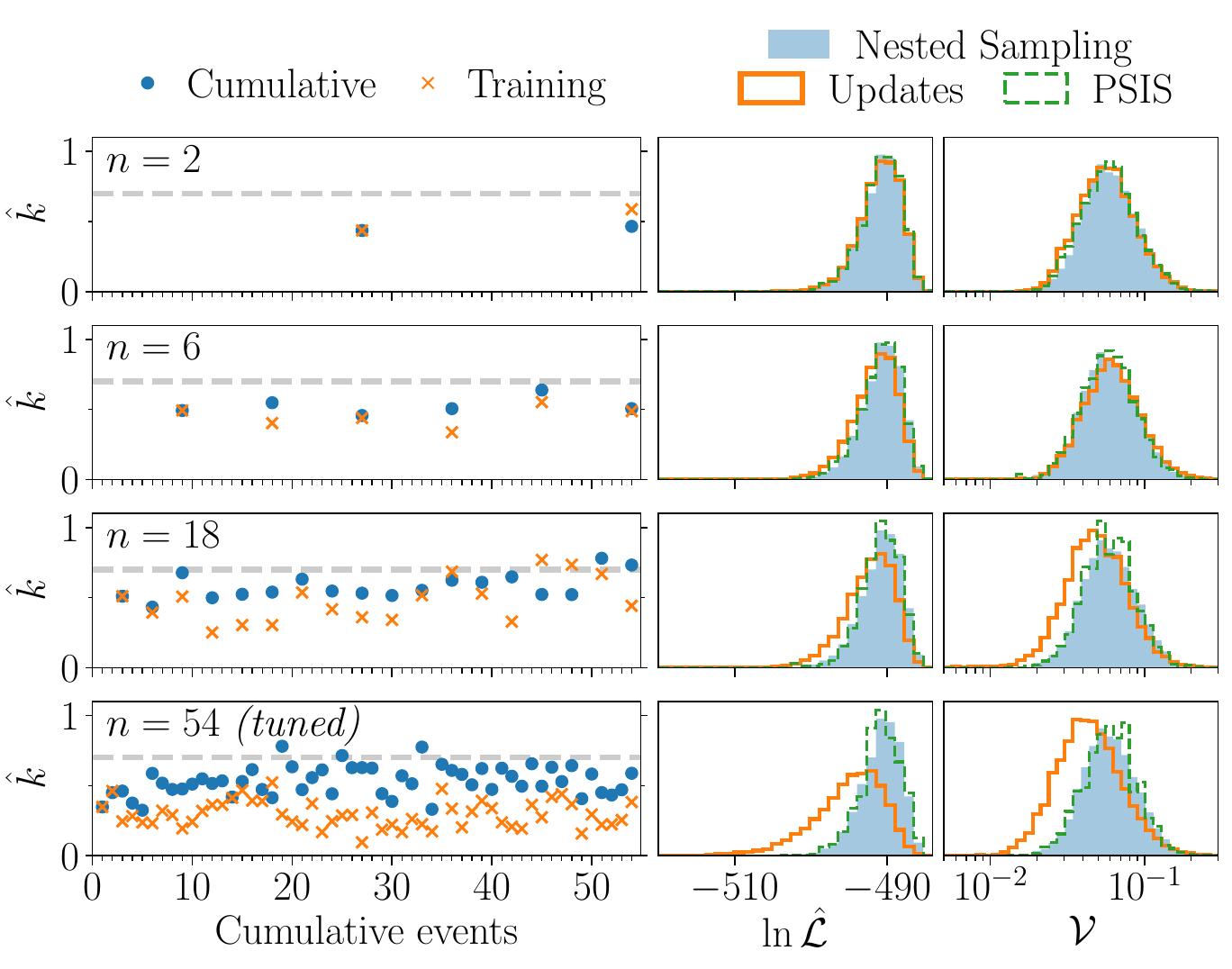}
    \caption{
    Convergence of sequential updates under a strongly-modeled approach
    with a mock catalog of high \ac{SNR} sources,
    as we decrease the number of events in each update (top to bottom).
    We also record the number of updates $n$ in each row.
    Format otherwise matches Fig.~\ref{fig:gwtc4-convergence}
    }
    \label{fig:strong-convergence}
\end{figure}

In the left column of Fig.~\ref{fig:strong-convergence}
we evaluate the convergence of $\mathcal{Q}_m$ after each update with $\hat{k}$;
we also show posterior distributions of $\ln \hat{\mathcal{L}}$ and variance $\mathcal{V}$ given the entire dataset, after the last update.
First, we find that performing updates with many, small updates
is less accurate.
When performing $n = 2$ or 6 updates,
the variational approximant is always converged to the training density
and cumulative posterior.
When performing $n = 18$ updates, however, we find that \ac{VI} sometimes
lies on the edge of convergence ($\hat{k} \gtrsim 0.7$) relative to the
training density.
In turn, $\hat{k}$ relative to the cumulative posterior increases
over update steps, which may indicate that errors in the variational approximant
to the posterior propagate forward as errors in the prior for the next update.
By lowering the learning rate when performing single-event updates,
we avoid some accumulation of error as single-event updates
remain converged ($\hat{k} < 0.7$) after most updates.
In general,
by inspecting the posterior distributions of $\ln \hat{\mathcal{L}}$ and $\mathcal{V}$
(right column of Fig.~\ref{fig:strong-convergence}),
we see that many sequential updates with small subcatalogs of data
tend to return a less accurate representation of the final, cumulative posterior,
with tails towards smaller log-likelihoods and larger variances than returned by nested sampling.
However, we are still able to recover the same posterior as nested sampling
with \ac{PSIS}.

\begin{figure}
    \centering
    \includegraphics[width=0.8\linewidth]{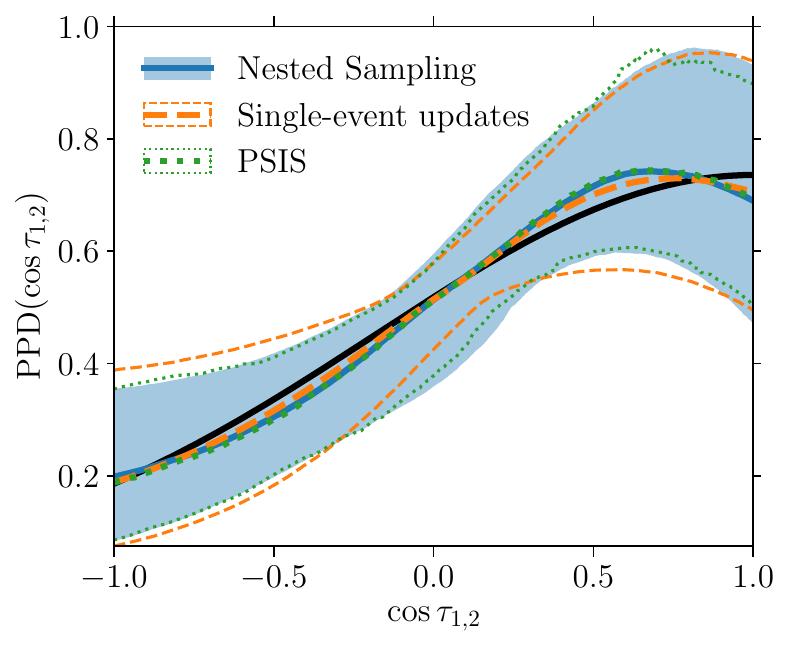}
    \caption{
    \acsp{PPD} in cosine-spin tilts $\cos \tau_{1,2}$
    returned by nested sampling (blue, filled),
    54 single-event updates (orange, dashed),
    and \acl{PSIS} of the variational approximant after all single-event updates
    (\ac{PSIS}, green).
    Thick lines denote the medians;
    filled regions or lines enclose the 90\% credible interval.
    A black line denotes the true distribution.
    }
    \label{fig:strong-ppd-cos-tau}
\end{figure}

Finally, in Fig.~\ref{fig:strong-ppd-cos-tau} we compare the \ac{PPD} in $\cos \tau_{1,2}$ returned by nested sampling
to that returned after $n = 54$ sequential updates
and \ac{PSIS} of the variational approximant after 54 updates.
We see that nested sampling and sequential updates roughly agree that there is a peak at $\cos \tau_{1, 2} \sim 0.8 -1$,
although \ac{PSIS} is required to recover the same median and 90\% credible region as nested sampling.
These results suggest that limiting our catalog to high \ac{SNR} sources where $\cos \tau_{1,2}$ are typically better measured ameliorates accumulated errors in the marginal spin tilt distribution.
We note, however, that the difficulty of the \ac{VI} task and the efficiency of importance sampling after each update will also be sensitive to the form and dimension of the population model;
here, we used a lower dimensional population model than in our analysis of GWTC-4 ($D = 15$ vs. $D = 19$).

\subsection{Weakly-modeled population} \label{sec:nb10}

We now test how these results depend on the model used for \ac{GW} population inference.
In particular, we consider a more flexible model that does not make strong assumptions about the functional form of the underlying population.
In principle, we could sequentially update a flexible model with real data,
although we anticipate similar challenges as with spin tilts in Sec.~\ref{sec:updates-o4a}
since most events on their own may provide relatively little information
about most of a flexible model's parameters.

Here, we model the joint space of primary mass $m_1$ and redshift $z$ with \textsc{PixelPop} \cite{Heinzel:2024jlc},
evenly dividing over $\ln m_1$ and $z$ in $N_b \times N_b$ bins and inferring the merger rate in each bin.
We adopt an \ac{ICAR} prior \cite{car, car-and-icar, Alvarez-Lopez:2025ltt}
that sets a log-normal prior on the merger rate in each bin, each with nearest-neighbor couplings to ensure a degree of smoothness of the merger rate while still allowing flexibly for correlations between mass and redshift.
We reuse the strong population models and priors
for $q$, $\chi_{1,2}$, and $\tau_{1,2}$ from Sec~\ref{sec:strong}.
Our population model can be written
\begin{align} \label{eq:pp}
    \mathcal{R}(\theta | \Lambda) = \mathcal{R}(\ln m_1; z | \{ \mathcal{R}_b \} ) p(q, \chi_{1,2}, \tau_{1,2} | m_1, \lambda) \, ,
\end{align}
where the differential merger rates $\{ \mathcal{R}_b \}$ in each bin $b$ and the parameters $\lambda$ of the strong, uncorrelated population model form the entire set of population model parameters $\Lambda$.
Note that $q$ is conditioned on $m_1$ as we taper the $q$ distribution at small secondary masses $m_2 = q m_1$.
We first test \ac{VI} and neural sequential updates with $N_b = 10$, giving $\Lambda$ a dimension of $D = 107$.
We space bins logarithmically over $m_1/M_\odot \in [3,150]$ and uniformly over $z \in [0, 0.5]$.
We also fix the minimum secondary \ac{BH} mass to $3 M_\odot$.

The \ac{ICAR} prior has one nuisance parameter, the coupling strength $\sigma$ between bins.
The population posterior under this prior is
$\mathcal{P}( \Lambda, \sigma | \mathcal{D} ) \propto \mathcal{L}(\mathcal{D} | \Lambda) \pi(\lambda) \pi(\{ \mathcal{R}_b \} | \sigma) \pi(\sigma)$.
Ultimately, we are only interested in the rate distribution implied by our data;
further, in the context of Bayesian updates, $\sigma$ only modifies later posteriors
as the prior density encoded in the first posterior.
Since the likelihood does not depend on $\sigma$,
to marginalize $\mathcal{P}$ over $\sigma$ we only need to marginalize
the \ac{ICAR} prior on the rates: $\pi ( \{ \mathcal{R}_b \} ) = \int \dd{\sigma} \pi ( \{ \mathcal{R}_b \} | \sigma ) \pi(\sigma)$.
During \ac{VI} or sequential updates,
we numerically marginalize the \ac{ICAR} prior over a uniform prior on $\ln \sigma \in [-3, 3]$. 

Before Bayesian updating, we determine whether \ac{VI} can fit a $107$-dimensional posterior on $\Lambda$.
With an initial learning rate $\eta = 10^{-1}$,
we experimented with increasing block dimensions of 8, 16, and 64
and found that a block dimension of 64 provided the most accurate fit to the posterior.
We train the variational approximant,
clipping gradients to a maximum norm of 1,
for $10^5$ steps with a batch size $B = 10$;
a total of $10^6$ likelihood evaluations performed over $20$\,minutes.
We achieve $\hat{k} = 0.69$ and $\varepsilon = 1.4$\%, indicating convergence.
For comparison, we perform \ac{HMC} using the No U-Turn Sampler \cite{2011arXiv1111.4246H}
implemented in \texttt{numpyro} \cite{phan2019composable, bingham2019pyro}.
We use $2.5 \times 10^5$ warmup steps, thinning post-warmup every 500 steps,
yielding 1010 posterior samples.
Warmup and sampling took approximately $2 \times 10^7$
likelihood evaluations,
performed over $20$\,hours.

\begin{figure}
    \centering
    \includegraphics[width=0.9\linewidth]{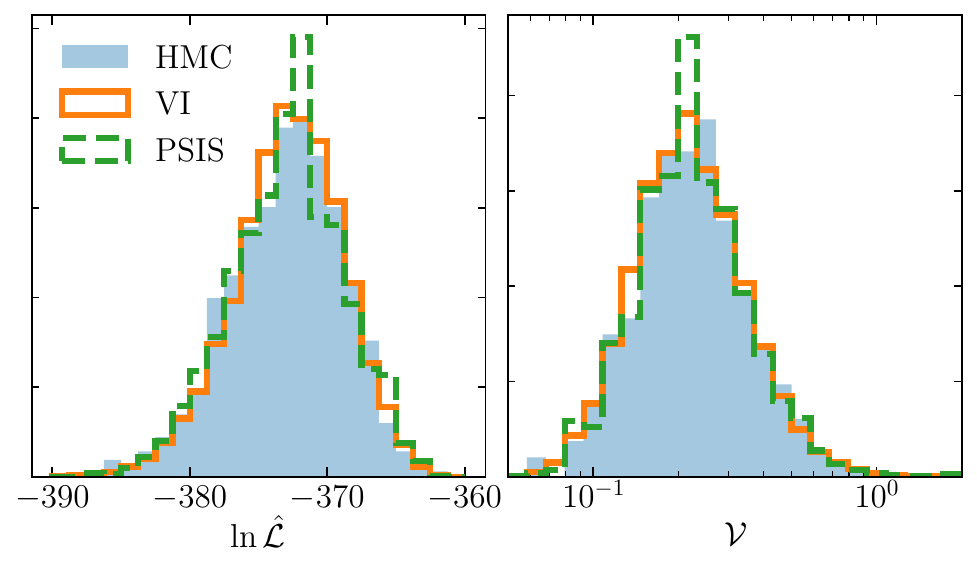}
    \caption{Posterior distributions of log-likelihood estimate $\ln \hat{\mathcal{L}}$ and variance $\mathcal{V}$ under a weakly-modeled approach with $10 \times 10$ bins in log-primary mass $m_1$ and redshift $z$.
    Results shown are obtained with \acl{HMC} (HMC, blue), \acl{VI} (VI, orange), and \acl{PSIS} of the variational approximant (PSIS, green).
    }
    \label{fig:nb10-lnl-var}
\end{figure}

In Fig.~\ref{fig:nb10-lnl-var}, we compare the posterior distributions of
the log-likelihood estimator and its variance
as achieved with \ac{HMC} and \ac{VI}.
These distributions appear indistinguishable within the finite number of posterior samples drawn using both methods
(although note a spuriously large weight which increases the density at the mode
when importance sampling; such weights can occur more often when importance
sampling in high-dimensional spaces \cite{2008arXiv0805.3034B}).

\begin{figure}[h]
    \centering
    \includegraphics[width=0.99\linewidth]{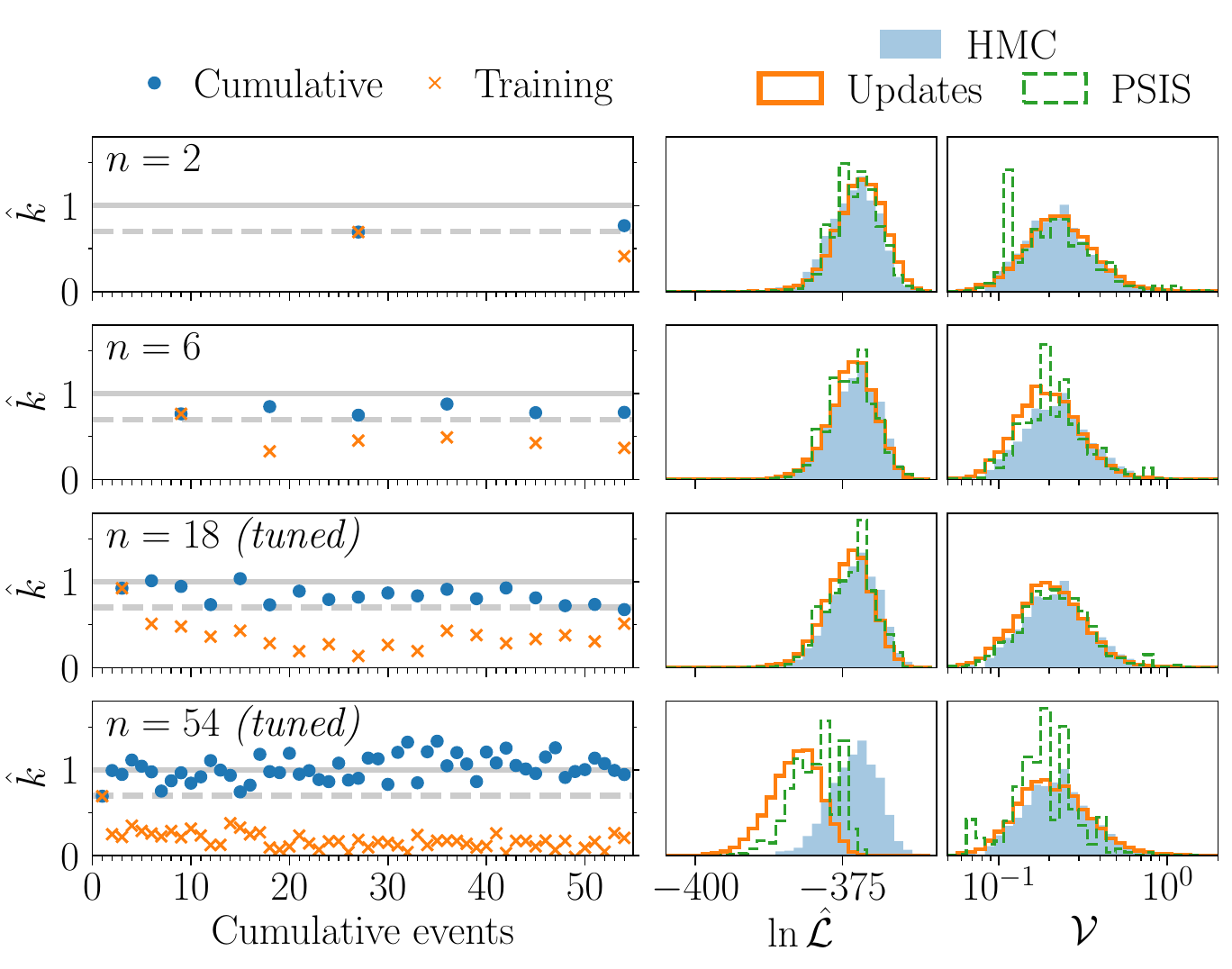}
    \caption{
    Convergence of sequential updates under a weakly-modeled approach.
    Format matches Fig.~\ref{fig:strong-convergence}.
    }
    \label{fig:nb10-convergence}
\end{figure}

We attempt to reproduce these results through a series of Bayesian updates,
dividing the mock catalog into subcatalogs as in Sec.~\ref{sec:strong}.
We first evaluate sequential updates with the same training settings
($\eta = 10^{-1}, B = 10, 10^5$\,steps)
as when performing \ac{VI} with the entire catalog.
In the top two rows of Fig.~\ref{fig:nb10-convergence},
we plot the convergence of $n = 2$ and 6 sequential updates with
these settings.
We see that $\hat{k} \lesssim 0.7$
relative to the training density
and $\hat{k} \sim 0.7$ relative to the cumulative density after each update.

When performing more updates with fewer events---$n = 18$ and 54 updates---we need to coarsely
tune the learning rate and flow size to achieve convergence.
We repeat each of these sequences of updates over a grid of $\eta \in \{ 10^{-1}, 10^{-2}, 10^{-3} \}$
and block dimension in $\{ 8, 16, 64 \}$.
We also test a constant learning rate of $10^{-3}$ at each block dimension.
Training with $B = 10$ and $10^5$ steps,
we find that the best results (lowest final $\hat{k}$ relative to the cumulative posterior)
using $\eta = 10^{-2}$ and a block dimension of 16,
yielding $\hat{k} \approx 0.86$ after $n = 18$ updates
and $\hat{k} \approx 0.95$ after $n = 54$ single-event updates.
We repeat $n = 18$ and 54 sequential updates while training for longer,
with $\eta = 10^{-2}$ and $10^6$ steps.
For $n = 18$ updates, $\eta = 10^{-2}$, $10^6$ steps, and a block dimension of 16
yielded the best final result, with $\hat{k} \approx 0.67$ and $\varepsilon \approx 1.5\%$.
For $n = 54$ updates, this same experiment yielded final convergence criteria of
$\hat{k} \approx 1.2$ and $\varepsilon \approx 0.052\%$.
Note that a cosine-decay learning schedule is also determined by the number of training
steps---training longer at fixed $\eta$ means more steps with
learning rates similar to the initial rate.
Therefore, we experiment with additional variations of $\eta$
while training for $10^5$ or $10^6$ steps,
although we still achieve the best single-event updating results with $\eta = 10^{-2}$ and $10^5$ steps.
All training configurations used for $n = 54$ updates are summarized
in Tab.~\ref{tab:single-event-update-experiments}.

\begin{table}[h]
    \begin{tabular}{ccccc}
    \toprule
    \toprule
    \multicolumn{2}{c}{Learning schedule} & \multicolumn{3}{c}{Block dimension} \\
    \cmidrule(lr){1-2} \cmidrule(lr){3-5}
    $\eta$ & $\eta_f$ & 8 & 16 & 64 \\
    \midrule
    $10^{-1}$ & 0         & 1.44 & 1.10 & 2.14 \\ 
    $10^{-2}$ & 0         & 1.08 & $\mathbf{0.95}$ & 1.11 \\
    $10^{-3}$ & 0         & 1.30 & 1.40 & 1.54 \\
    $10^{-3}$ & $10^{-3}$ & 1.60 & 1.62 & 1.60 \\
    $10^{-4}$ & 0         & ---  & 1.19 & --- \\
    $10^{-5}$ & 0         & ---  & 1.54 & --- \\
    \midrule
    $10^{-2}$ & 0         & ---  & 1.19 & --- \\
    $10^{-3}$ & 0         & ---  & 1.61 & --- \\
    $10^{-4}$ & 0         & ---  & 1.47 & --- \\
    $10^{-5}$ & 0         & ---  & 1.11 & --- \\
    \bottomrule
    \bottomrule
    \end{tabular} 
    \caption{
    Pareto-shape $\hat{k}$ of the importance weights after the final
    single-event update under a weakly-modeled approach
    achieved with different learning schedules
    (specified by initial $\eta$ and final $\eta_f$ learning rates)
    and block dimension of the variational family.
    The learning schedule is also determined by the number of steps;
    either $10^5$ (top) or $10^6$ (bottom).
    Block dimensions of 8, 16, and 64 correspond to variational families with
    $1.9 \times 10^5$, $3.7 \times 10^5$, and $1.5 \times 10^6$ trainable parameters, respectively.
    The smallest $\hat{k}$ is bolded.
    }
    \label{tab:single-event-update-experiments}
\end{table}

We evaluate the convergence of these tuned $n = 18$ and 54 updates
in the bottom two rows of Fig.~\ref{fig:nb10-convergence}.
For $n = 2$, 6, and 18 updates, we successfully recover the same posteriors on
$\ln \hat{\mathcal{L}}$ and $\mathcal{V}$ as \ac{HMC},
albeit with noticeable importance sampling noise.
This follows as the final Pareto shape in all cases,
relative to the cumulative posterior, is $\hat{k} \gtrsim 0.7$,
on the edge of convergence.
Nominally, this might warn us that the variational approximant is not converged to the true posterior.
However, it is only when we attempt $n = 54$ updates that we fail to recover the posterior
distribution of $\ln \hat{\mathcal{L}}$; here, $\hat{k} \sim 1$.
In practice, importance weight ratios grow rapidly with dimension,
making it challenging to achieve convergence in high-dimensional spaces \cite{2015arXiv150702646V}.
Therefore, while $\hat{k} < 0.7$ may be sufficient for convergence,
our variational approximants with $0.7 < \hat{k} \lesssim 1$
may still be astrophysically useful
(see Refs.~\cite{2019arXiv191004102H, Uzsoy:2024zho} for further discussion). 

We also notice that $\hat{k} \gtrsim 0.7$ after the first update at all $n$,
and in particular $\hat{k} \lesssim 1$ after the first of $n = 18$ updates.
While the Pareto shape may be an overly conservative convergence indicator
in high dimension,
we attempted to improve convergence by changing the initialization of the first
variational approximant $\mathcal{Q}_1$.
When fitting the posterior given the first event
we initialized $\mathcal{Q}_1$
by fitting it to an ``effective prior'',
$\pi( \{ \mathcal{R}_b \}, \lambda ) \times \exp(-\sum_{b}^{N_b} \mathcal{R}_b^2)$.
While this initialization did not improve the convergence of the first
single-event update,
the choice of effective prior is not unique.
We leave the exploration of better \ac{VI} initialization strategies
targeting densities with improper priors to future work.

\begin{figure*}
    \centering
    \includegraphics[width=0.9\linewidth]{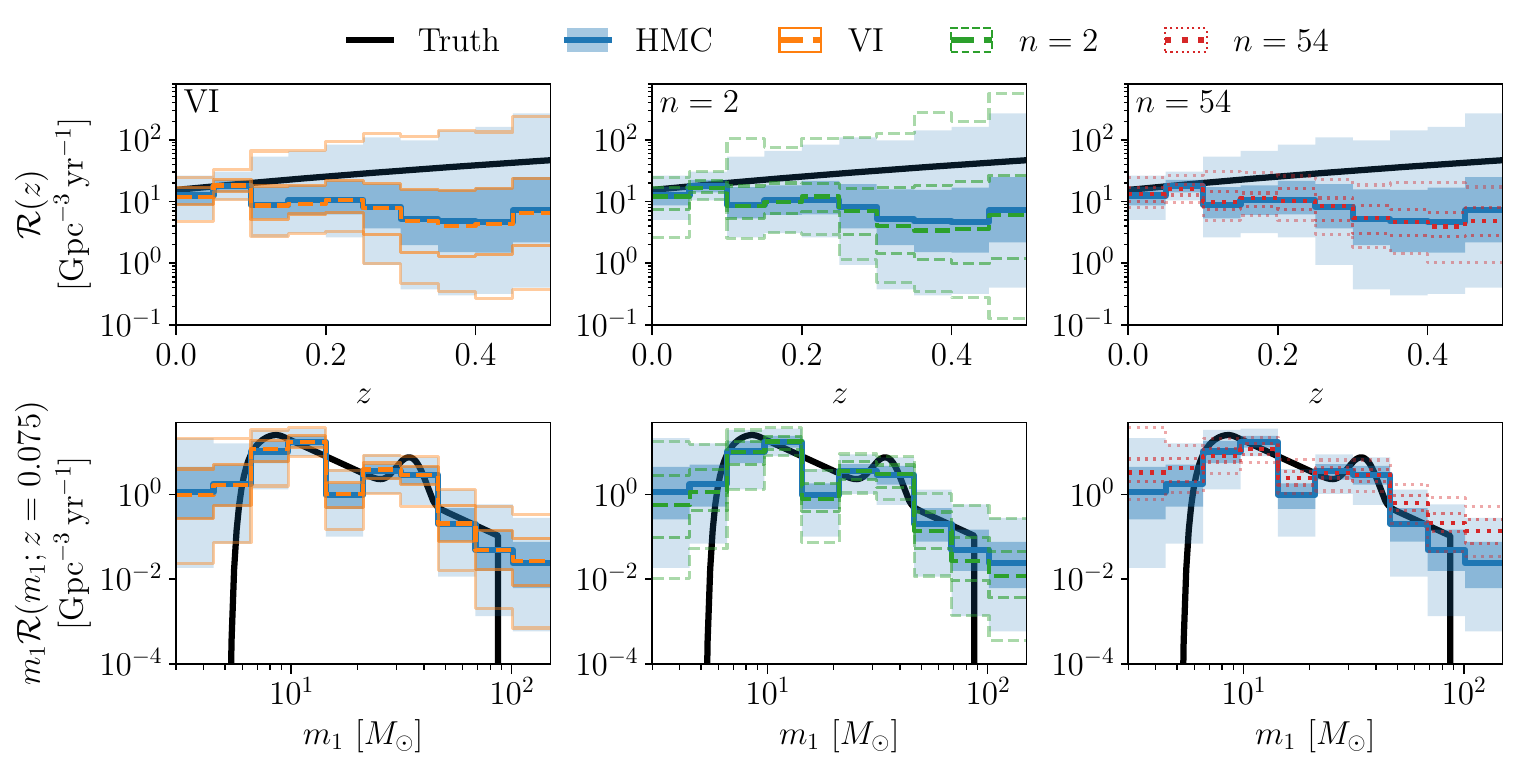}
    \caption{
    Posterior rate distributions in redshift $z$ (top)
    and log primary mass $m_1$ (bottom)
    under a weakly-modeled approach with $10 \times 10$ bins;
    as reported by \acl{HMC} (HMC; blue, filled) and
    \acl{VI} (VI; left column),
    $n = 2$ sequential updates (middle column),
    and $n = 54$ single-event updates (right column).
    Posteriors for \ac{VI} or updates are shown after \ac{PSIS}.
    Light (dark) lines or shading denote the 50\% (90\%) credible region.
    A solid (broken) line shows the median reported by \ac{HMC} (\ac{VI} or updates).
    The true astrophysical distribution is shown in black.
    }
    \label{fig:all-ppds}
\end{figure*}

To understand where the variational approximant after sequential updates succeeds and fails,
in Fig.~\ref{fig:all-ppds} we inspect the
posterior differential rate distributions in $z$ and $\ln m_1$
(at $z = 0.075$, where $\mathcal{R}(z)$ is best measured)
as returned by \ac{VI}, $n = 2$, and $n = 54$ updates.
We find that with only 10 bins along each axis
the model does not capture the astrophysical $m_1$ distribution within 90\%
credibility near $m_1 \sim 13\,M_\odot$
(we repeated inference using 40 bins with \ac{HMC} and recovered the true distribution at 90\% credibility).
We also note that the merger rate is consistently overestimated at $m_1 \lesssim 5\,M_\odot$
and $m_1 \gtrsim 100\,M_\odot$, a known effect of the \ac{ICAR} prior \cite{Alvarez-Lopez:2025ltt}.

Crucially, however, these are failures of our coarse population model---which \ac{VI} captures with similar certainty as \ac{HMC} at all masses and redshifts.
The posterior returned by $n = 2$ updates agrees qualitatively with \ac{HMC},
returning a similar shape for the $\ln m_1$ and $z$ distributions,
although the variational approximant noticeably underestimates the median,
50\% credible bounds, and lower 90\% credible bound
where the marginal merger rate is more poorly measured
(at $m_1 \lesssim 7\,M_\odot$, $m_1 \gtrsim 40\,M_\odot$, and $m_1 \sim 13\,M_\odot$; see Fig.~\ref{fig:all-ppds}, middle column).
Finally, $n = 54$ single-event updates successfully recovers the median
rate in redshift---marginalized over $m_1$---but otherwise dramatically
underestimates the width of the 50\% and 90\% credible regions (see Fig.~\ref{fig:all-ppds}, right column).
Single-event updating only recovers a similar median marginal rate in $\ln m_1$ as \ac{HMC}
in those bins where $\mathcal{R}(\ln m_1; z=0.075)$ is best measured;
everywhere, single-event updating fails to recover similar 50\% and 90\%
credible bounds on the marginal rate in mass.
This may be an example of ``mode-seeking'',
where the variational approximant fails to capture the full posterior support, 
which is a general feature of \ac{VI} using the form of the \ac{KL} divergence
in Eq.~\eqref{eq:kl} \cite{bishop2006pattern}.
When performing variational sequential updates,
mode-seeking may compound as an artificially-narrow variational posterior
becomes a narrow prior when analyzing new data.

\begin{figure}[htb!]
    \centering
    \includegraphics[width=0.8\linewidth]{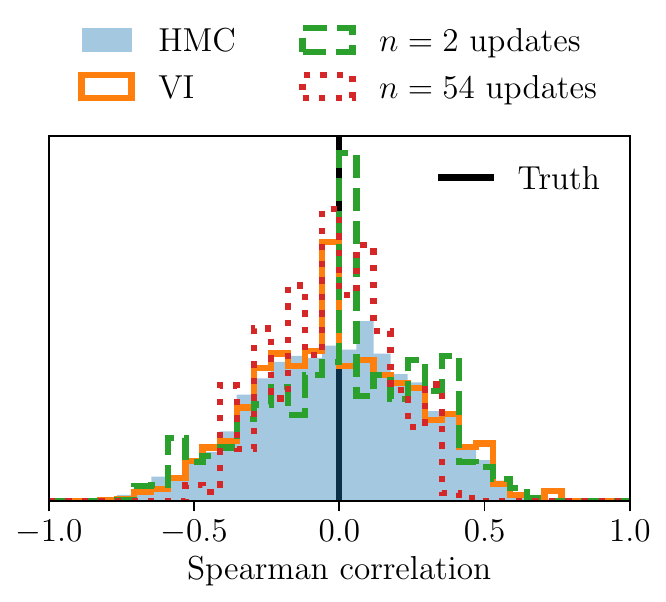}
    \caption{
    Posteriors on the Spearman rank coefficient quantifying monotonic correlations between (log) primary mass and redshift under the weakly-modeled approach with $10 \times 10$ bins,
    as reported by
    \acl{HMC} (HMC, blue, filled),
    \acl{VI} (VI, orange),
    $n = 2$ updates (green, dashed),
    and $n = 54$ single-event updates (red, dotted). 
    The truth---no correlation---is shown with a solid black line.
    }
    \label{fig:nb10-spearman}
\end{figure}

Next, we investigate whether variational inference and sequential updates
can capture correlations between $\ln m_1$ and $z$.
We compute the Spearman rank correlation coefficient \cite{Spearman1904, 1979ats..book.....K, schweizer1981nonparametric} 
following the method in \citet{Heinzel:2024jlc}.
In Fig.~\ref{fig:nb10-spearman}, we compare posteriors on the Spearman coefficient
returned by \ac{HMC}, \ac{VI}, $n = 2$ updates,
and $n = 54$ single-event updates.
We find that the \ac{HMC}, \ac{VI}, and $n = 2$ sequential updates posteriors are effectively the same,
while single-event updating returns a more narrow posterior,
with a 90\% credible interval $40$\% more narrow than \ac{HMC}.
The variational posteriors after both $n = 2$ and $n = 54$ updates
suffer noticeable importance sampling noise (cf. Fig.~\ref{fig:nb10-convergence}).
In one sense, Fig.~\ref{fig:nb10-spearman} is encouraging
because \ac{VI} and sequential updates
reach similar astrophysical conclusions as \ac{HMC};
namely, that we cannot confidently claim there are, or are not, mass--redshift correlations
in the underlying population.
On the other hand, compounding mode-seeking over many sequential updates
may tend to miss correlations in the high-dimensional posterior 
and prefer no correlation regardless of the underlying distribution or
the observed data.

\section{Conclusions} \label{sec:conclusions}

Gravitational-wave catalogs are quickly growing larger and more complex.
Analysts seek to fit increasingly flexible, high-dimensional models to these data
to identify the formation channels of merging \ac{BBH}.
Luckily, Bayesian statistics provides a natural framework for breaking down complex \ac{GW} datasets:
sequential updates of population priors as new events arrive,
instead of re-analyzing the entire cumulative catalog.
Sequential updates of \ac{GW} population models could enable
analyses with larger models and datasets on finite-memory hardware accelerators,
and during an observing run analysts could:
update \ac{GW} event rate estimates;
update astrophysically-informed significance estimates (e.g., $p_{\rm astro}$);
and inform astrophysical population model development ahead of public data releases. 
In this work,
we leveraged rapid posterior density estimation enabled by neural \ac{VI} \cite{Mould:2025dts}
to put sequential updates into practice.
We addressed the following questions--

\textit{Can we perform sequential updates with neural \acl{VI}?}:
We first tested sequential updates with real \ac{GW} catalogs,
adopting a strong model from the \ac{LVK} analysis of \acp{BBH} in \ac{GWTC}-4 \cite{LIGOScientific:2025pvj}.
Starting from a variational fit to the \ac{BBH} in \ac{GWTC}-3,
we updated the population with \acp{BBH} observed during O4a at three cadences:
with all O4a events at once,
month-by-month,
and as each event arrived during the observing run.
In each case,
sequential updates produced variational approximants
that could be importance sampled to recover similar distributions
in primary mass, mass ratio, spin magnitude, and redshift as nested sampling.
Single-event updating necessitated some tuning of the learning schedule to achieve convergence.
Sequential updates failed to recover the same spin-tilt distribution,
with single-event and then monthly updates yielding the least
accurate distributions.
We conclude that sequential updates can be performed in a
strongly-modeled approach to the masses, redshift, and spin magnitude distributions of real \ac{BBH}
populations,
while the spin tilt distribution appears the most susceptible to variance---both Poisson and numerical---and
requires greater scrutiny.
In our experiments,
we sought to reproduce typical population analyses that consider all events at once.
We carefully accounted for selection effects among different observing periods.
However, sequential updates during an observing run would instead use approximations
to the detector sensitivity which may not be exact as, e.g.,
detector sensitivity can change on monthly or weekly timescales \cite{Essick:2025zed, Essick:2025bna};
we leave a detailed study of population updates with approximate
sensitivity estimates to future work.

\textit{Can we extend neural sequential updating to future catalogs?}:
We next performed sequential updates with a small mock catalog of high-\ac{SNR}
sources,
taken from a catalog comparable in size to the number of sources
expected by the end of the \acp{LVK} next observing run.
When performing updates with the current public catalog,
we noted that performing more updates with fewer events
degraded the accuracy of the final variational spin-tilt posterior---which may
be related to typically poor measurements of source spin tilts.
To test this hypothesis with our mock catalog,
we used a relatively high \ac{SNR} threshold of 30.
We again took a strongly-modeled approach to the population,
adopting a low-dimensional, phenomenological model from \citet{KAGRA:2021duu}.
Performing many sequential updates with few observations tended to yield a less
accurate fit to the posterior given the cumulative dataset.
However, by reducing the learning rate and number of variational family parameters,
we were able to importance sample the final variational approximant
to the same posterior density as nested sampling---even for spin tilts,
and even when performing single-event updates.

\textit{Can we extend neural \acl{VI} and sequential updating to high-dimensional population models?}:
Next, we took a weakly-modeled approach,
employing \textsc{PixelPop} \cite{Heinzel:2024jlc} to model the distribution of
log primary mass and redshift in $10 \times 10$ evenly spaced, correlated bins.
We used this population model to again analyze our mock catalog of high \ac{SNR} sources.
We successfully recovered the $D = 107$-dimensional population posterior
with \ac{VI} in $20$ minutes compared to $20$ hours with \ac{HMC}.
We also successfully recovered this high-dimensional posterior distribution
when performing 2, 6, and 18 sequential updates
(with 27, 9, and 3 events at a time, respectively).
Just as with sequential updates under strong population models,
we found that using fewer parameters
to specify the variational family and reducing the initial learning rate
improved the convergence of the approximant after many updates with few events.
This implies one possible explanation to why error accumulates over $n$ sequential updates:
we are attempting to train a factor of $n$ additional variational parameters
with the same finite information provided by the cumulative catalog of \ac{GW} sources.
Also, since we transfer information between updates by initializing 
each approximant with the best-fit parameters from the previous update,
reducing the initial learning rate may balance regularization
of the Monte Carlo estimate of the loss enabled by large learning rates
\cite{2017arXiv170807120S} without immediately scattering the variational
parameters far from their initialization.
While these insights improved the convergence of our variational approximant
when updating with 3 events or 1 event at a time;
the variational posterior after 54 single-event updates
failed to reproduce marginal distributions in mass and redshift.
Further tuning of the learning schedule or other training settings
may limit mode-seeking when performing single-event updates
of high-dimensional population models.

\textit{What other science is enabled by neural sequential updating?}:
One immediate application of neural sequential updating is evaluating information gain
from \ac{BBH} events or subcatalogs of events,
since we have a tractable density estimator after each update.
This provides another alternative to leave-one-out analyses for quantifying
the impact of each event on the inferred population
without modeling the ``full'' hierarchical likelihood
\cite{Mancarella:2025uat}.
We briefly outlined this application with GW231123,
an event inferred to have extremal masses and spins \cite{LIGOScientific:2025rsn},
and which we noted---with sequential updating---pulled the inferred population
towards broader spin magnitude distributions.
We emphasize, however, that this statement is conditioned on the events
observed prior to GW231123,
not the entire catalog at the very end of the observing run.

Neural updates to \ac{GW} population models also yield
an efficient representation of the population likelihood
which can be combined with other analyses and searches as new data arrive;
for example, with constraints on the stochastic background of unresolved mergers to infer high-redshift \ac{BBH} evolution \cite{Callister:2020arv, Bers:2025tei},
with electromagnetic counterparts to compact object mergers to measure
cosmological parameters \cite{2005ApJ...629...15H, Salvarese:2025qel}
or the speed of gravity \cite{LIGOScientific:2017zic, Iampieri:2024dul},
or with rates of gamma-ray bursts to constrain the nuclear equation of state \cite{Chen:2025otp}.
Sequential updating, as presented here, is also naturally applicable to
spectral siren cosmological inference \cite{Chernoff:1993th, Taylor:2011fs, LIGOScientific:2025jau}
and tests of general relativity \cite{LIGOScientific:2021sio, Magee:2023muf}
which can be formulated as generalizations of astrophysical population models.
Although in this work we used the population likelihood marginalized over source parameters,
as we showed in Sec.~\ref{sec:updates} Bayesian updates can be performed with the joint likelihood as well;
neural updates of the full hierarchical likelihood could provide real-time,
astrophysically-informed constraints on the observability of counterparts
for neutron star--black hole mergers,
which are sensitive to the mass and spin of the source compact objects
(see \citet{Biscoveanu:2022iue} and references therein).
Beyond \ac{GW} population analyses,
neural sequential updating can be applied to
Bayesian searches for stochastic gravitational-wave backgrounds,
where the likelihood factorizes in time between data segments
as in the inhomogeneous Poisson likelihood we used \cite{Smith:2017vfk, Smith:2020lkj, Biscoveanu:2020gds}.
Similarly,
searches for continuous \ac{GW} sources
have already been rephrased in terms of Bayesian updating,
whether updating parameter estimation of suspected continuous wave sources as in \citet{LIGOScientific:2013rhu},
or combining searches over smaller segments of data in the manner of
hierarchical ``semi-coherent'' searches
(see \citet{Tenorio:2021wmz} for a review and \citet{Martins:2025jnq} which has performed Bayesian updates between search stages, as opposed to sequential data segments).
Sequential updating is also a natural solution to global-fit type problems \cite{Cornish:2005qw}.

In total,
neural sequential updating under strong and weak assumptions
about the \ac{BBH} population can yield useful inferred population distributions.
We explored how model complexity and the information content in each update
modify the accuracy of sequential updating.
We found that tuning the training procedure and the variational family
improved the accuracy of variational approximations to the posterior,
even for high-dimensional population models and over many updates with few events.
We have not exhausted the space of training configurations;
one immediate variation to consider are other variational families
besides the block-neural autoregressive flow used
in this work, particularly for even higher-dimensional posteriors
which may tend to be approximately Gaussian.
Other applications could consider tuning the learning procedure based
on the anticipated information gain from an update e.g. the \ac{SNR} of
a new \ac{GW} event when updating population analyses.
With these lessons in hand,
our work demonstrates that neural sequential updating is a useful technique
for analyzing growing datasets delievered by current and next-generation gravitational-wave observatories.

\acknowledgments

We thank Tom Callister, Deep Chatterjee, Storm Colloms, Berthy Feng, Maya Fishbach, Jack Heinzel, Colm Talbot, and Dominika Zięba for discussions,
and Max Isi for internal LIGO review.
N.E.W. is supported by the National Science Foundation Graduate Research Fellowship Program under grant No. 2141064.
M.M. is supported by a Royal Commission for the Exhibition of 1851 Research Fellowship and by the LIGO Laboratory through the National Science Foundation award No. PHY-2309200.
S.V. is partially supported by the NSF grant No. PHY-2045740.
JV is supported by STFC grants ST/V005634/1 and UKRI2487.
The authors are grateful for computational resources provided by
the LIGO Laboratory supported by National Science Foundation Grants PHY-0757058 and PHY-0823459.
This material is based upon work supported by NSF's LIGO Laboratory which is a major facility fully funded by the National Science Foundation and has made use of data or software obtained from the Gravitational Wave Open Science Center (gwosc.org), a service of the LIGO Scientific Collaboration, the Virgo Collaboration, and KAGRA.
This work is supported by the National Science Foundation under Cooperative Agreement PHY-2019786 (The NSF AI Institute for Artificial Intelligence and Fundamental Interactions, http://iaifi.org/).

\appendix

\section{Priors} \label{app:priors}

\begin{table}[h!]
    \centering
    \begin{tabular}{ p{5 cm} c c }
        \toprule
        \toprule
        \textbf{Parameter} & \textbf{Symbol} & \textbf{Prior} \\
        \midrule
        $m_1$ low-mass power law index & $\alpha_1$ & $[-4, 12]$ \\
        $m_1$ high-mass power law index & $\alpha_2$ & $[-4, 12]$ \\
        Break mass & $m_{\rm break}$ & $[20, 50]\,M_\odot$ \\
        First peak location & $m_{1}$ & $[5, 20]\,M_\odot$ \\
        First peak width & $\sigma_{1}$ & $[0, 10]\,M_\odot$ \\
        Second peak location & $m_{2}$ & $[25, 60]\,M_\odot$ \\
        Second peak width & $\sigma_{2}$ & $[0, 10]\,M_\odot$ \\
        Low-mass smoothing scale & $\delta_m$ & $[0, 10]\,M_\odot$ \\
        Minimum \ac{BH} mass & $m_{\min{}}$ & $[3, 5.5]\,M_\odot$ \\
        Mixing fractions between broken power law and peaks & $\lambda_0, \lambda_1$ & Dirichlet \\
        $q$ power law index & $\beta$ & $[-2, 7]$ \\
        \midrule
        Mean & $\mu_\chi$ & $[0, 1]$ \\
        Variance & $\sigma_\chi^2$ & $[0.005, 1]$ \\
        \midrule
        Peak location & $\mu_\tau$ & $[-1, 1]$ \\
        Peak width & $\sigma_\tau$ & $[0.01, 4]$ \\
        Peak mixing fraction & $\xi_\tau$ & $[0, 1]$ \\
        \midrule
        $z$ power law index & $\lambda_z$ & $[-10, 10]$ \\
        Log total merger rate & $\ln K / \mathrm{yr}$ & $[8, 12]$ \\
        \bottomrule
        \bottomrule
    \end{tabular}
    \caption{
    Priors when analyzing the \ac{BBH} in \ac{GWTC}-4.
    These parameters specify the astrophysical distribution of
    masses (top),
    spin magnitudes (second from top),
    spin tilts (second from bottom),
    and redshifts (bottom).
    We denote a uniform prior between $a$ and $b$ as $[a, b]$.
    The Dirichlet prior is equally-weighted between all components.
    The total merger rate $K$ is measured in the frame of the detector.
    }
    \label{tab:gwtc-priors}
\end{table}

\begin{table}[h!]
    \centering
    \begin{tabular}{ c c c }
        \toprule
        \toprule
        \textbf{Parameter} & \textbf{Symbol} & \textbf{Prior} \\
        \midrule
        $m_1$ power law index & $\alpha$ & $[1.5, 5]$ \\
        $q$ power law index & $\beta$ & $[3, 7]$ \\
        Peak mean & $\mu_m$ & $[26, 39]\,M_\odot$ \\
        Peak standard deviation & $\sigma_m$ & $[1, 11]\,M_\odot$ \\
        Peak mixing fraction & $\lambda_m$ & $[0, 0.5]$ \\
        Minimum \ac{BH} mass & $m_{\min{}}$ & $[3, 7]\,M_\odot$ \\
        Maximum \ac{BH} mass & $m_{\max{}}$ & $[80, 100]\,M_\odot$ \\
        Low-mass smoothing scale & $\delta_m$ & $[0, 15]\,M_\odot$ \\
        High-mass smoothing scale & $\delta_{\max}$ & $[0, 10]\,M_\odot$ \\
        \midrule
        Shape alpha & $\alpha_\chi$ & $[0, 10]$ \\
        Shape beta & $\beta_\chi$ & $[0, 10]$ \\
        \midrule
        Peak standard deviation & $\sigma_\tau$ & $[0.2, 2]$ \\
        Peak mixing fraction & $\xi_\tau$ & $[0, 1]$ \\
        \midrule
        $z$ power law index & $\lambda_z$ & $[-9, 10]$ \\
        Log total merger rate & $\ln K / \mathrm{yr}$ & $[2, 14]$ \\
        \bottomrule
        \bottomrule
    \end{tabular}
    \caption{
    Priors on the parameters of strongly-modeled population distributions used when analyzing the mock catalog in Sec.~\ref{sec:mock}; format matches Tab.~\ref{tab:gwtc-priors}.
    The beta distribution we adopt for spin magnitudes is nonsingular,
    so the alpha and beta parameters of that distribution are greater than one.
    }
    \label{tab:strong-priors}
\end{table}

In Tab.~\ref{tab:gwtc-priors},
we show priors on the parameters of
strongly-modeled population distributions used to analyze \ac{GWTC}-4
(Sec.~\ref{sec:gwtc4}).
In Tab.~\ref{tab:strong-priors},
we show priors on the parameters of different strongly-modeled population
distributions used to analyze a mock catalog (Sec.~\ref{sec:mock}).

\section{Number of synthetic signals in O4a} \label{app:ninj}

In Sec.~\ref{sec:gwtc4}, we perform sequential updates of an astrophysical population model with subsets of the data observed during O4a.
As part of the update likelihood,
we estimate the number of sources expected in each batch of data
via weighted Monte Carlo sum over synthetic signals added to detector data
(cf. Sec.~\ref{sec:monte-carlo}).
Reweighting these signals to a uniform prior in detector-frame time, following Ref.~\cite{2021RNAAS...5..220E},
requires the total number of synthetic signals produced over a given duration.
The \ac{LVK} has produced sensitivity estimates for synthetic signals in distinct sets for O1, O2, O3, and each month of O4a \cite{Essick:2025zed, o4a-sensitivity-zenodo, gwtc4-cumulative-sensitivity-zenodo}.
The number of signals and total detector-frame time for O1, O2, and O3 are already included in public data releases;
in Tab.~\ref{tab:ninj}, courtesy of \citet{ninj},
we record the size and duration of synthetic signal sets for each month of O4a.
Note that we exclude synthetic signals generated during the engineering run (ER15)
prior to O4a;
thus we compute the duration of the synthetic signal set in May 2023 from the official start of O4a, at GPS time 1368975618 \cite{LIGOScientific:2025slb}.

\begin{table}[h]
    \centering
    \begin{tabular}{ c c c c }
        \toprule
        \toprule
        Month (MM/YY) & Start [s] & End [s] & Total generated \\
        \midrule
        05/23 & 1366933504 & 1369612288 & 60499005 \\
        06/23 & 1369612288 & 1372205056 & 53692138 \\
        07/23 & 1372205056 & 1374882304 & 53318599 \\
        08/23 & 1374882304 & 1377561088 & 49285598 \\
        09/23 & 1377561088 & 1380153856 & 44134007 \\
        10/23 & 1380153856 & 1382831104 & 43618343 \\
        11/23 & 1382831104 & 1385423872 & 41870421 \\
        12/23 & 1385423872 & 1388102656 & 43738263 \\
        01/24 & 1388102656 & 1390779904 & 43001343 \\
        \bottomrule
        \bottomrule
    \end{tabular}
    \caption{Total number of synthetic signals (detected and non-detected)
    generated for each month of O4a. 
    The start and end of each month are provided in GPS time.
    All values courtesy of \citet{ninj}.
    }
    \label{tab:ninj}
\end{table}

\section{Approximate cumulative log-likelihood variance} \label{app:var-approx}

\begin{figure}
    \centering
    \includegraphics[width=0.85\linewidth]{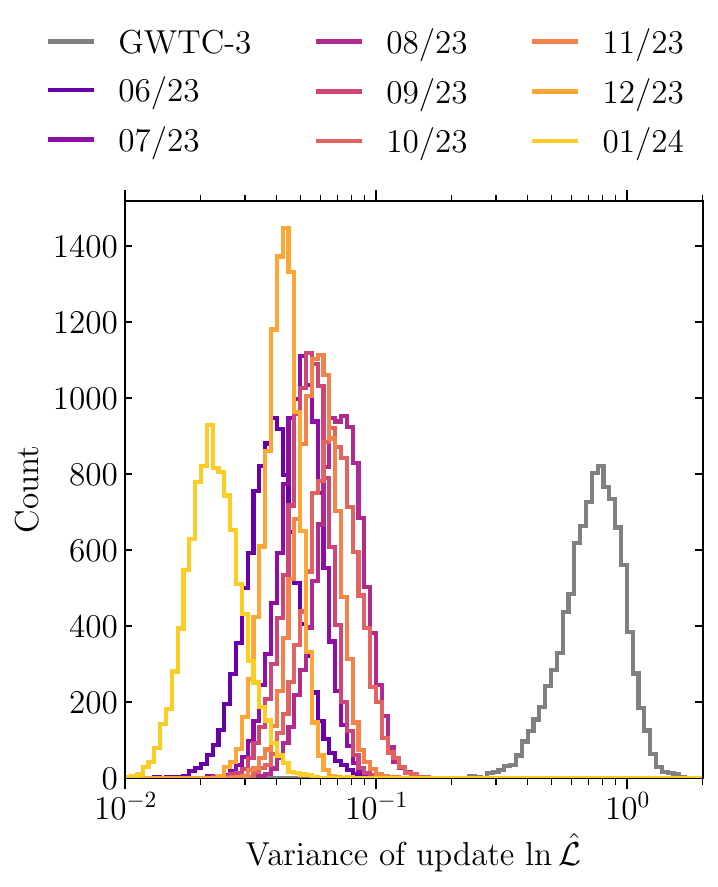}
    \caption{Variance of the log-likelihood estimator for each month of data
    during O4a (colors) and \ac{GWTC}-3 (grey).
    Variances are computed over the posterior support after each monthly update;
    posterior samples have \textit{not} been importance sampled
    to the cumulative posterior.
    }
    \label{fig:monthly-variances}
\end{figure}

\begin{figure}
    \centering
    \includegraphics[width=0.65\linewidth]{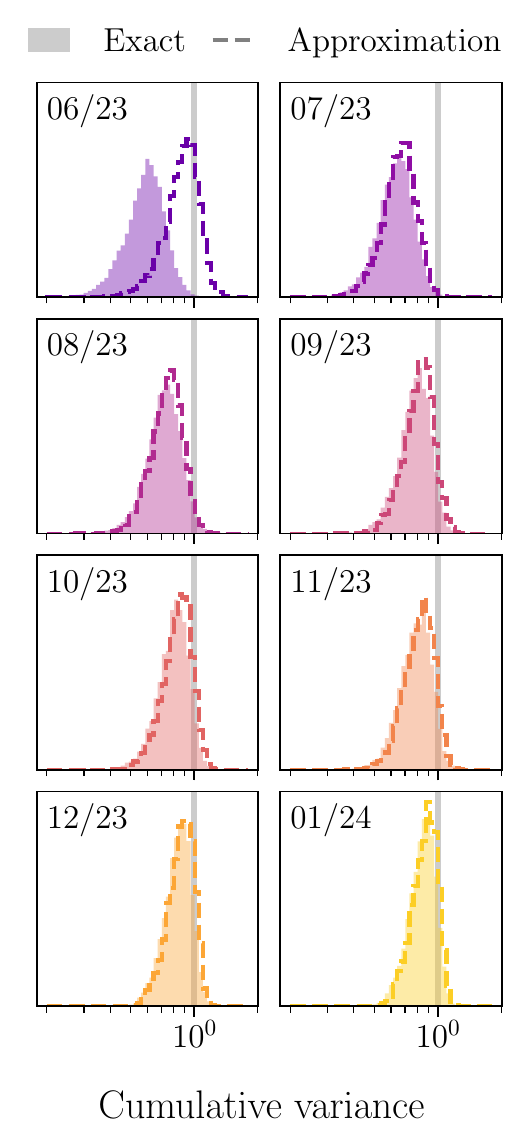}
    \caption{Variance of the cumulative log-likelihood estimator
    after each month of O4a (filled) and the upper bound on the variance
    constructed according to Eq.~\eqref{eq:var-decomposed} and approximated
    with a neural network (dashed).
    }
    \label{fig:monthly-approx}
\end{figure}

When performing population inference with Monte Carlo estimates of the
likelihood,
analyses sometimes exclude regions of the population parameter $\Lambda$ space
with uncertain estimates of the log-likelihood \cite{LIGOScientific:2025pvj, Tiwari:2017ndi, 2019RNAAS...3...66F, Essick:2022ojx, Talbot:2023pex, jack_math_tome}.
When performing sequential updates,
we find that the variance of the log-likelihood for a single update
can be much smaller than the cumulative log-likelihood.
In Fig.~\ref{fig:monthly-variances}, we show the variance
of the log-likelihood estimator for each month of observations during O4a.
We observe that the variance of the update log-likelihood is typically
one to two orders of magnitude smaller than the cumulative variance,
of the log-likelihood of either GWTC-3 or GWTC-4
(cf. Fig.~\ref{fig:gwtc4-convergence}).
Therefore, tapering on the variance of the update log-likelihood estimator
may be insufficient to regularize estimates the cumulative likelihood.
One straightforward solution is to compute and taper on the variance of the
cumulative log-likelihood estimate while performing sequential updates.
The primary downside of this approach is that computational cost
of performing $n$ updates will scale like $n (n + 1) / 2 \sim n^2$.
Alternatively, we could use additional single-event posterior samples,
since we downsample each posterior to the minimum sample size across events.
While this would make update and cumulative log-likelihood estimates
more accurate,
if we do not track the cumulative log-likelihood estimate during each update
it may be possible to recover population posteriors with support
where the update likelihood estimates are relatively certain
but the cumulative log-likelihood is highly uncertain.
Instead, here we construct an approximate upper bound on the cumulative
variance which use to regularize sequential updates with real data in Sec.~\ref{sec:gwtc4}.

We can write the variance of the cumulative log-likelihood estimator as
\begin{equation}
    \mathcal{V}(\Lambda) = \sum_{i = 1}^{N} v_i(\Lambda) + (\rnorm{} T)^2 v_{\rm det}(\Lambda)
\end{equation}
where $v_i$ is the variance of the Monte Carlo estimate of the log of Eq.~\ref{eq:marg-single-event-like}
and $v_{\det}$ is the variance of the log of the fraction $\alpha_{\det}$
of detectable sources;
assuming constant detector sensitivity,
$\alpha_{\det} = \bar{N} K^{-1} T^{-1}$ where $\bar{N}$ is the number of expected
sources from Eq.~\eqref{eq:nexp} and $K$ is the total detector-frame \ac{BBH}
merger rate from Eq.~\eqref{eq:model-explicit-rate}.
Dividing our catalog into two periods of observation as in Sec.~\ref{sec:updates},
the variance can be written,
\begin{equation} \label{eq:var-start}
\begin{aligned}
    \mathcal{V} &= \sum_{i = 1}^{N_1} v_i + \sum_{j = 1}^{N_2} v_j + (K (T_1 + T_2))^2 v_{\rm det} \\
    &= \sum_{i = 1}^{N_1} v_i + \sum_{j = 1}^{N_2} v_j + K^2(T_1^2 + T_2^2 + 2 T_1 T_2) v_{\rm det} \, \\
\end{aligned}
\end{equation}
where $N_{(1,2)}$ and $T_{(1,2)}$ are the number of observations and duration
of each period of observation.
Here, $v_{\rm det}$ is computed using all of the (detectable) simulated sources
added to detector data in both observing periods.
A Monte Carlo estimate of $\ln \alpha_{\det}$ over the first period of observation---so using only simulated sources over that first period---will have a variance $v_{\det, 1}$.
Similarly, a Monte Carlo estimate of $\ln \alpha_{\det}$ over the second
period will have a variance $v_{\det, 2}$.
We expect that $v_{\det, (1,2)} > v_{\det}$ since the total number
of simulations used for Monte Carlo estimation adds between periods of observation.
In turn, $(T_1 + T_2)^2 v_{\det} < T_1^2 v_{\det, 1} + T_2^2 v_{\det, 2} + 2 T_1 T_2 v_{\det, 2}$,
which allows us to bound the cumulative variance in Eq.~\eqref{eq:var-start},
\begin{equation} \label{eq:var-decomposed}
    \mathcal{V} < \mathcal{V}_1 + \mathcal{V}_2 + 2 K^2 T_1 T_2 v_{\det, 2} \, ,
\end{equation}
where we define $\mathcal{V}_{(1,2)} = \sum_{i = 1}^{N_{(1,2)}} v_i + K^2 T_{(1,2)}^2 v_{\det, (1,2)}$.

In the context of sequential updates,
we can take $\mathcal{V}_1$ as the variance of the log-likelihood of all the data observed
so far,
and $\mathcal{V}_2$ as the variance of the log-likelihood of a new dataset
with which we are updating the posterior on $\Lambda$.
After each sequential update, we already calculate the
cumulative log-likelihood variance over $10^4$ samples drawn from the
variational approximant to evaluate its accuracy.
We can re-use these calculations to train an emulator
to the cumulative variance when performing the next update;
in particular, by emulating $\mathcal{V}_1$
we can efficiently estimate an upper bound on
$\mathcal{V}$ using Eq.~\ref{eq:var-decomposed}
while updating with the second period of observation.

To this end, after each update we train a neural network to predict 
the cumulative variance as a function of $\Lambda$,
implemented in \texttt{equinox} \cite{kidger2021equinox}.
The neural network has 2 layers, each 10 wide, with an exponential linear unit
activation \cite{2015arXiv151107289C} between each layer.
We split our samples 75\%-25\% into training and validation sets.
We train the neural network by minimizing the mean-square error with respect
to the cumulative variance over batches of 100 random samples from the training set.
We use a cosine-decay learning rate \cite{2017arXiv170807120S}
from an initial learning rate of $10^{-1}$ to zero over $10^4$ steps;
training takes $\sim 2 - 5$ seconds.

During the subsequent update we regularize Monte Carlo
estimates of the cumulative log-likelihood
by enforcing a taper on the upper bound of the cumulative variance
constructed according to Eq.~\eqref{eq:var-decomposed}
and approximated with a neural network.
In Fig.~\ref{fig:monthly-approx}, we compare the cumulative log-likelihood
variance with our approximate upper bound;
we see the most discrepancy when we update from \ac{GWTC}-3 with
the first month of data.
This is likely because the detector sensitivity increased between O3b and O4a
\cite{LIGOScientific:2025hdt}.

In Sec.~\ref{sec:gwtc4} we adopted a variance threshold of 1,
which is already a sufficient but not necessary requirement for unbiased
population inference with Monte Carlo estimates of the likelihood
\cite{Essick:2022ojx, jack_math_tome}.
Regularizing the likelihood according to
an upper bound on the variance is even more conservative.
Future applications of sequential updates to \ac{GW} population analyses
could benefit from alternative methods for
reducing Monte Carlo variance,
such as approximations to the selection function
\cite{Talbot:2020oeu, Wong:2020wvd, Gerosa:2020pgy, Mould:2022ccw, Chapman-Bird:2022tvu, Callister:2024qyq, Lorenzo-Medina:2024opt}
or fitting a density estimator to single-event posterior samples
\cite{Mould:2022ccw, Wysocki:2018mpo, Golomb:2021tll, Delfavero:2022pnq, Mancarella:2025uat, Bers:2025tei, Dax:2021tsq, 2024arXiv240302443K, Liu:2025snl, Hussain:2025llf, Guttman:2025jkv}.

\bibliography{refs}

\end{document}